\documentclass[prb,preprint,amsfonts,superscriptaddress,english,endfloats*,showpacs]{revtex4-1}
\usepackage{dcolumn}
\usepackage{amsmath}
\usepackage{amssymb}
\usepackage{graphicx}
\usepackage{bm}
\usepackage{color}
\usepackage[english]{babel}

\begin{document}

\graphicspath{ {figures/} }

\title{Probing the coupling between a doublon excitation and the charge-density wave in TaS$_2$ by ultrafast optical spectroscopy}

\vspace{2cm}

\author{Andreas Mann}
	\affiliation{Laboratory for Ultrafast Microscopy and Electron Scattering, IPHYS, \'{E}cole Polytechnique F\'{e}d\'{e}rale de Lausanne, CH-1015 Lausanne, Switzerland}

\author{Edoardo Baldini}
	\affiliation{Laboratory for Ultrafast Microscopy and Electron Scattering, IPHYS, \'{E}cole Polytechnique F\'{e}d\'{e}rale de Lausanne, CH-1015 Lausanne, Switzerland}
	\affiliation{Laboratory of Ultrafast Spectroscopy, ISIC, \'{E}cole Polytechnique F\'{e}d\'{e}rale de Lausanne, CH-1015 Lausanne, Switzerland}
	
\author{Ahmad Odeh}
	\affiliation{Laboratory of Ultrafast Spectroscopy, ISIC, \'{E}cole Polytechnique F\'{e}d\'{e}rale de Lausanne, CH-1015 Lausanne, Switzerland}

\author{Arnaud Magrez}
	\affiliation{Crystal Growth Facility, \'{E}cole Polytechnique F\'{e}d\'{e}rale de Lausanne (EPFL), CH-1015 Lausanne, Switzerland}

\author{Helmuth Berger}
	\affiliation{Crystal Growth Facility, \'{E}cole Polytechnique F\'{e}d\'{e}rale de Lausanne (EPFL), CH-1015 Lausanne, Switzerland}

\author{Fabrizio Carbone}
	\affiliation{Laboratory for Ultrafast Microscopy and Electron Scattering, IPHYS, \'{E}cole Polytechnique F\'{e}d\'{e}rale de Lausanne, CH-1015 Lausanne, Switzerland}

\date{\today}

\begin{abstract}
Recently, the switching between the different charge-ordered phases of 1$T$-TaS$_2$ has been probed by ultrafast techniques, revealing unexpected phenomena such as ``hidden'' metastable states and peculiar photoexcited charge patterns. 
Here, we apply broadband pump-probe spectroscopy with varying excitation energy to study the ultrafast optical properties of 1$T$-TaS$_2$ in the visible regime. 
By scanning the excitation energy in the near-IR region we unravel the coupling between different charge excitations and the low-lying charge-density wave state. 
We find that the amplitude mode of the charge-density wave exhibits strong coupling to a long-lived doublon state that is photoinduced in the center of the star-shaped charge-ordered Ta clusters by the near-IR optical excitation. 
\end{abstract}


\maketitle

\section{Introduction}

Tantalum disulfide (TaS$_2$) belongs to the class of transition metal dichalcogenides (TMDCs), most of which are layered, quasi-two-dimensional materials whose properties are governed by strong electron-electron (el-el) correlation and electron-phonon (el-ph) interaction. 
They exhibit diverse electronic and phononic phenomena~\cite{Rossnagel2011,Sipos2008,Dean2011,Devereaux2007,Chhowalla2013,Klemm2015}, most notably a variety of charge-density wave (CDW) and superconducting phases, and often show complex phase diagrams. 
In TMDCs, the transition metal atoms generally arrange in planes with a trigonal structure, and each of these planes is sandwiched by layers of chalcogen atoms with the same trigonal structure. 
The interlayer bonding is given by weak van der Waals-type forces, leading to reduced dimensionality. 
Besides applications as transistors, solar cells, and photodetectors~\cite{Gibney2015}, TMDCs are popular in the study of strong electronic correlations due to the high tunability of their properties by intercalation~\cite{Klemm2015} or external parameters like pressure~\cite{Sipos2008} and electric field~\cite{Stojchevska2014,Ma2016}. 

In this paper we study the octrahedrally coordinated 1$T$-polytype of TaS$_2$ (denoted TaS$_2$ in the following) that shows various CDW phases. 
Below 550~K, TaS$_2$ enters an incommensurate CDW (i-CDW) phase, in which the charge ordering in the Ta plane is not commensurate to any lattice vector. 
Cooling below 350~K, the CDW becomes nearly commensurate (n-CDW phase) and finally, below 180~K, commensurate (c-CDW phase). 
The commensurate CDW takes on the shape of clusters of 13 Ta atoms forming a regular pattern of ``David's star'' units and shows a multitude of phonon modes related to the CDW~\cite{Gasparov2002}. 

The structural changes are accompanied by strong changes of the electronic structure due to the reconstruction of the Brillouin zone~\cite{Smith1985,Rossnagel2006}: 
The twelve Ta 5$d$ electrons from the outer atoms of the star split evenly into two separate bands below the Fermi energy, as observed by angle-resolved photoemission spectroscopy (ARPES) and inverse ARPES (ARIPES) experiments~\cite{Arita2004,Sato2014}. 
The remaining electron from the central atom of the star is located in a narrow band around the Fermi level, separated from the other bands due to spin-orbit coupling~\cite{Rossnagel2006}. 
This band experiences Mott localization, and the material enters an insulating state~\cite{Fazekas1979}, which implies the importance of el-el correlations in the formation of the CDW. 
The central cluster atom’s electron was observed to be localized by scanning tunneling spectroscopy~\cite{Kim1994}. 
The ultrafast collapse of the Mott gap upon laser excitation seen in time-resolved ARPES experiments~\cite{Perfetti2006,Perfetti2008,Hellmann2010} was suggested to lead to a state of polaronic conductivity~\cite{Dean2011}, and can be reproduced by density functional theory (DFT) calculations~\cite{Shen2014}. 
While there is consensus about the Mott insulating character of TaS$_2$, there is still controversy about the origin of the CDW state. 
Suggested candidates include interlayer coupling~\cite{Ma2016,Bovet2003}, orbital ordering~\cite{Ritschel2015}, and the particular $k$-dependence of the el-ph coupling~\cite{Faraggi2016}. 
Renewed interest in the properties of TaS$_2$ was sparked by the discovery that various long-lived, metastable phases can be photoinduced by exploiting the strong interaction between the charge-order patterns and the band structure changes resulting from the associated lattice distortions~\cite{Stojchevska2014,Laulhe2015}. 

Here, we reveal microscopic information about the electronic states coupled to the CDW by investigating the interplay between the CDW amplitude mode and high-energy electronic excitations of 1$T$-TaS$_2$ using ultrafast broadband pump-probe spectroscopy. 
We coherently excite the amplitude mode by tuning the optical excitation in the near-IR region while measuring the material's transient reflectivity $\Delta R/R$ in the visible regime. 
The choice of non-overlapping excitation and detection energies allows us to access the off-diagonal terms of the two-dimensional energy space that contain information on the coupling between the CDW amplitude and charge excitations on the Mott-Hubbard energy scale. 
The resonant behavior of the amplitude mode exhibits two peaks associated to charge excitations inside the CDW star pattern that create doublon-holon pairs. 
We find that the CDW amplitude mode couples strongly to the doublon at the star's center, and that this doublon state is long-lived, persisting for several picoseconds. 

\begin{figure*}[tb]
	\includegraphics[width=.25\linewidth]{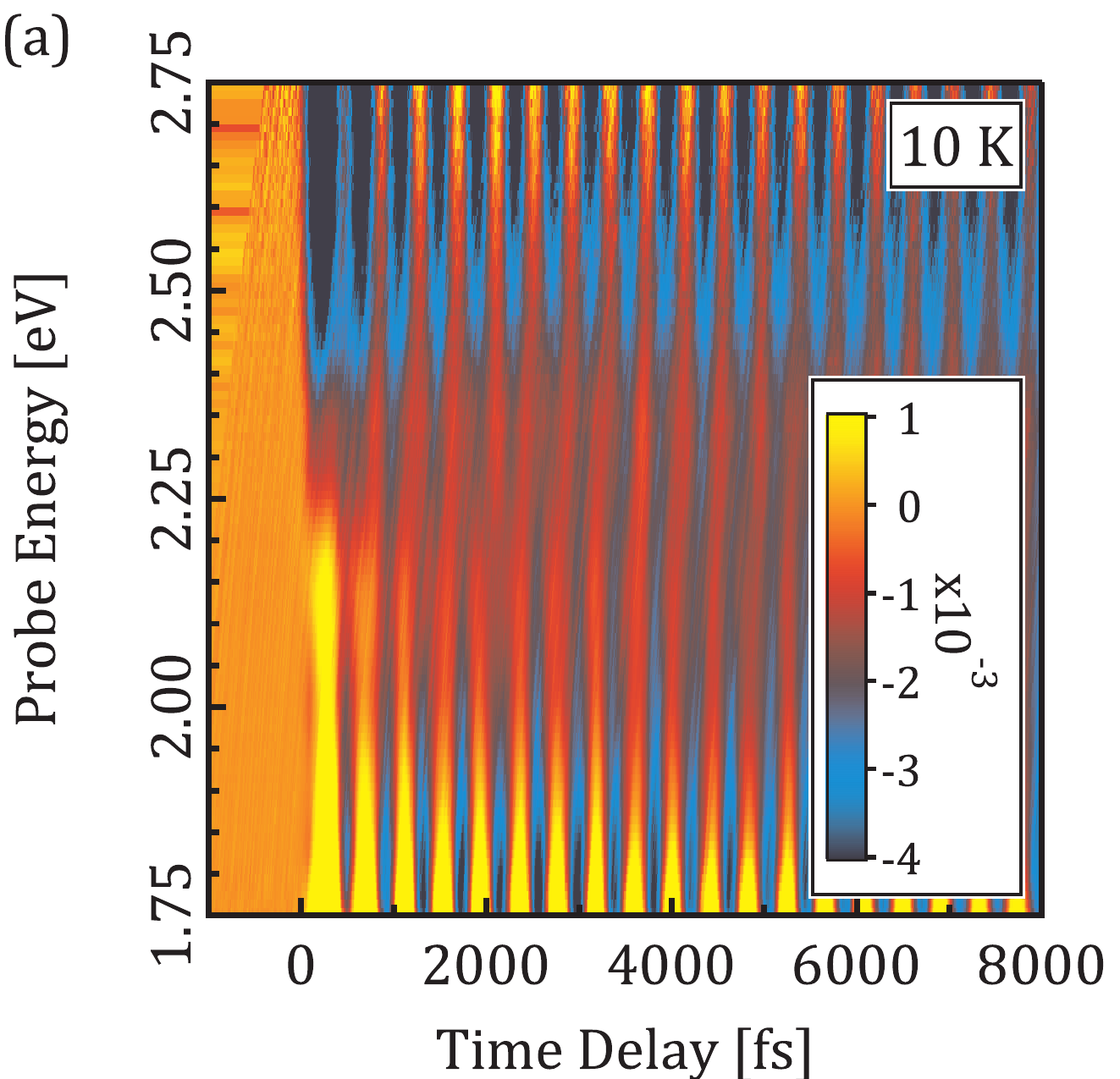}\hfill
	\includegraphics[width=.25\linewidth]{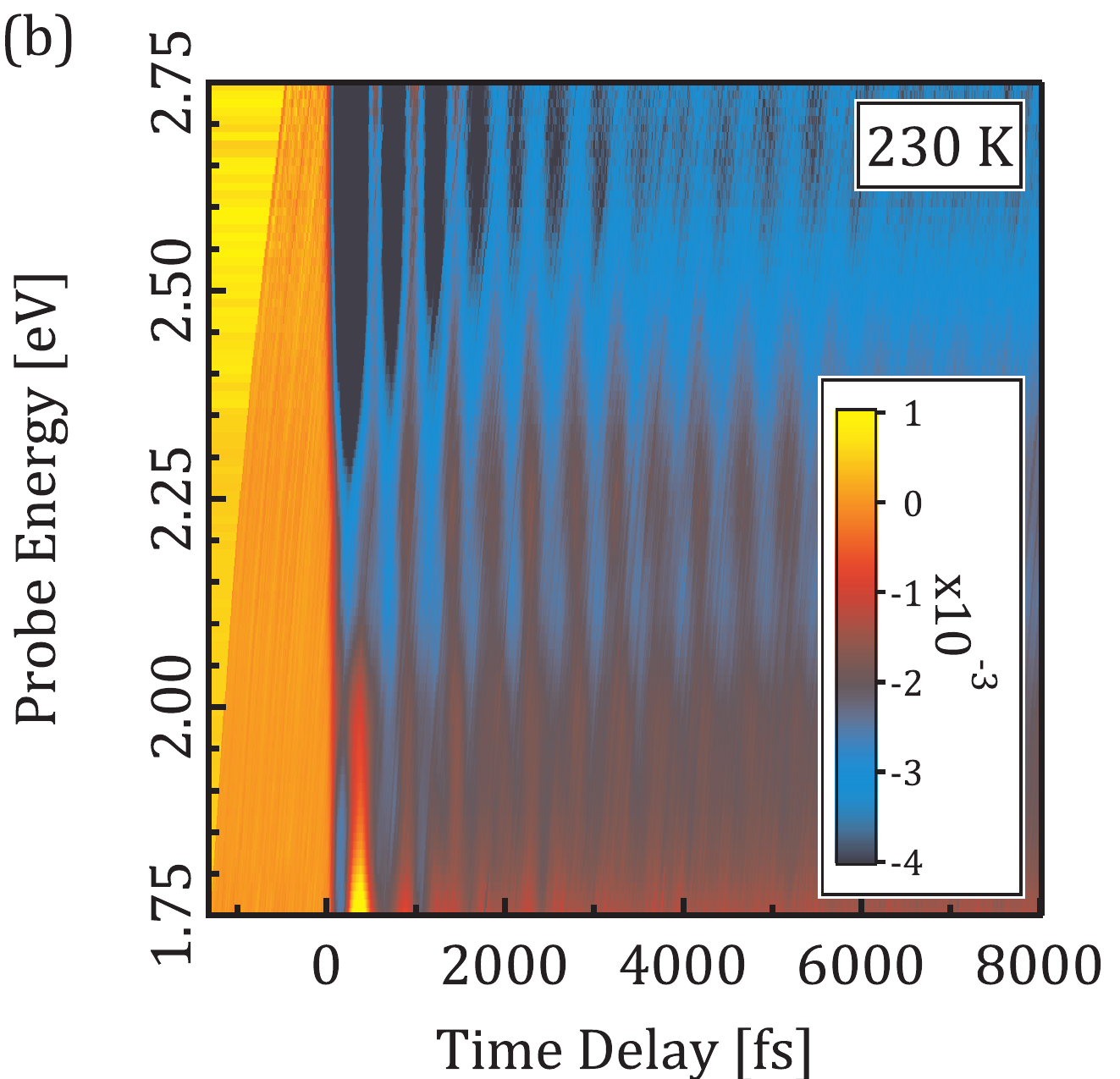}\hfill
	\includegraphics[width=.25\linewidth]{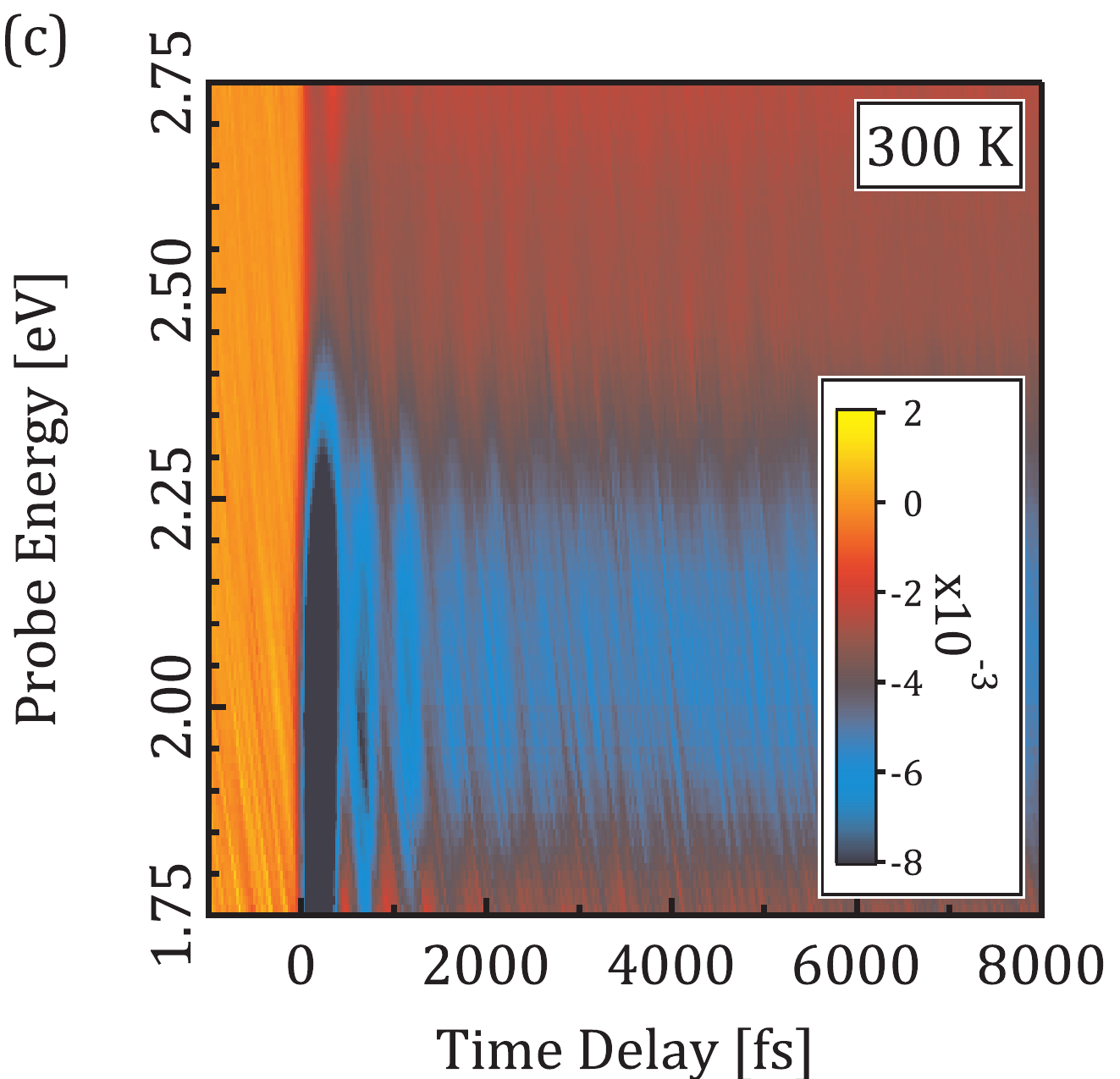}\hfill
	\includegraphics[width=.25\linewidth]{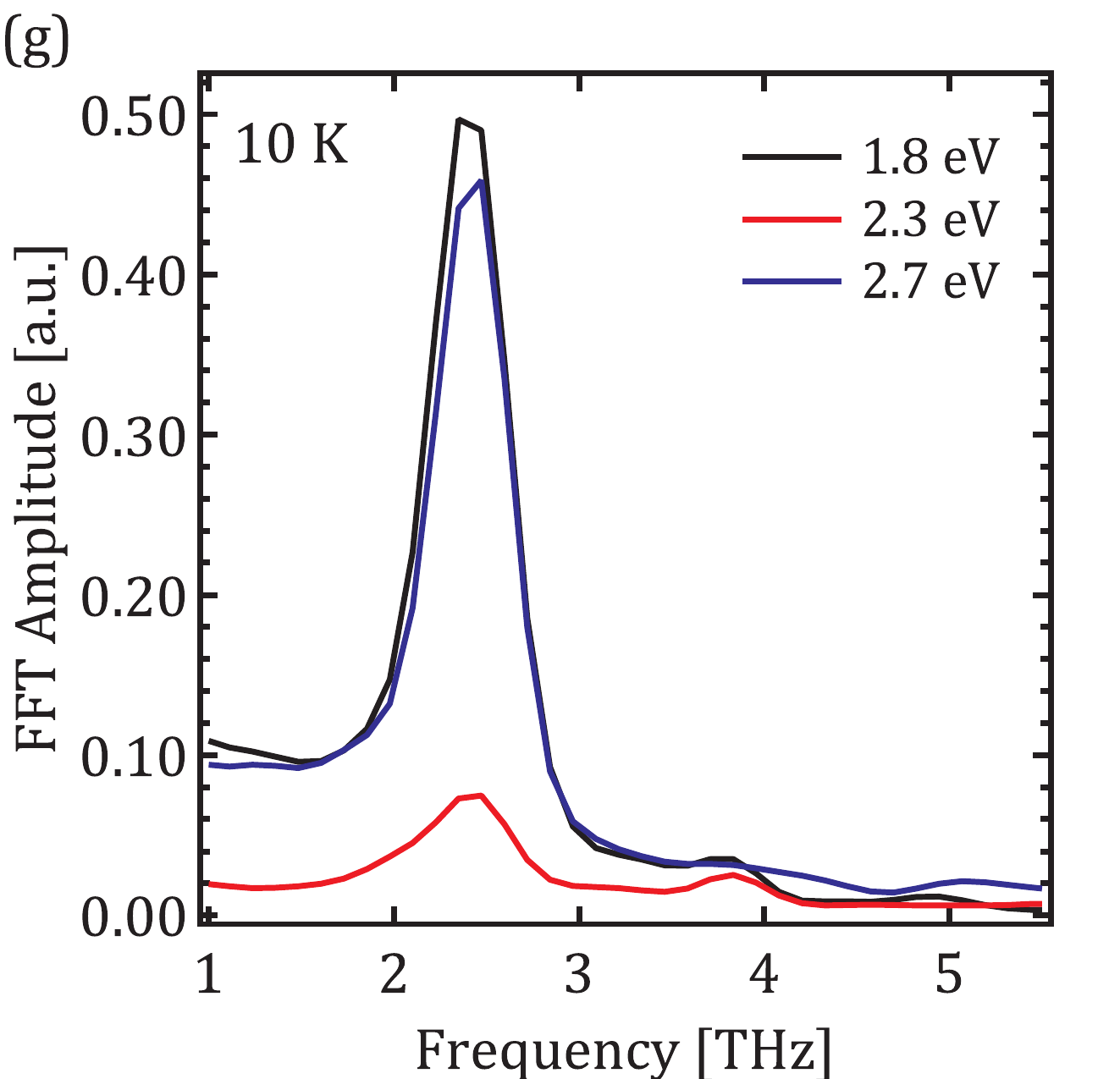}\\
	\includegraphics[width=.25\linewidth]{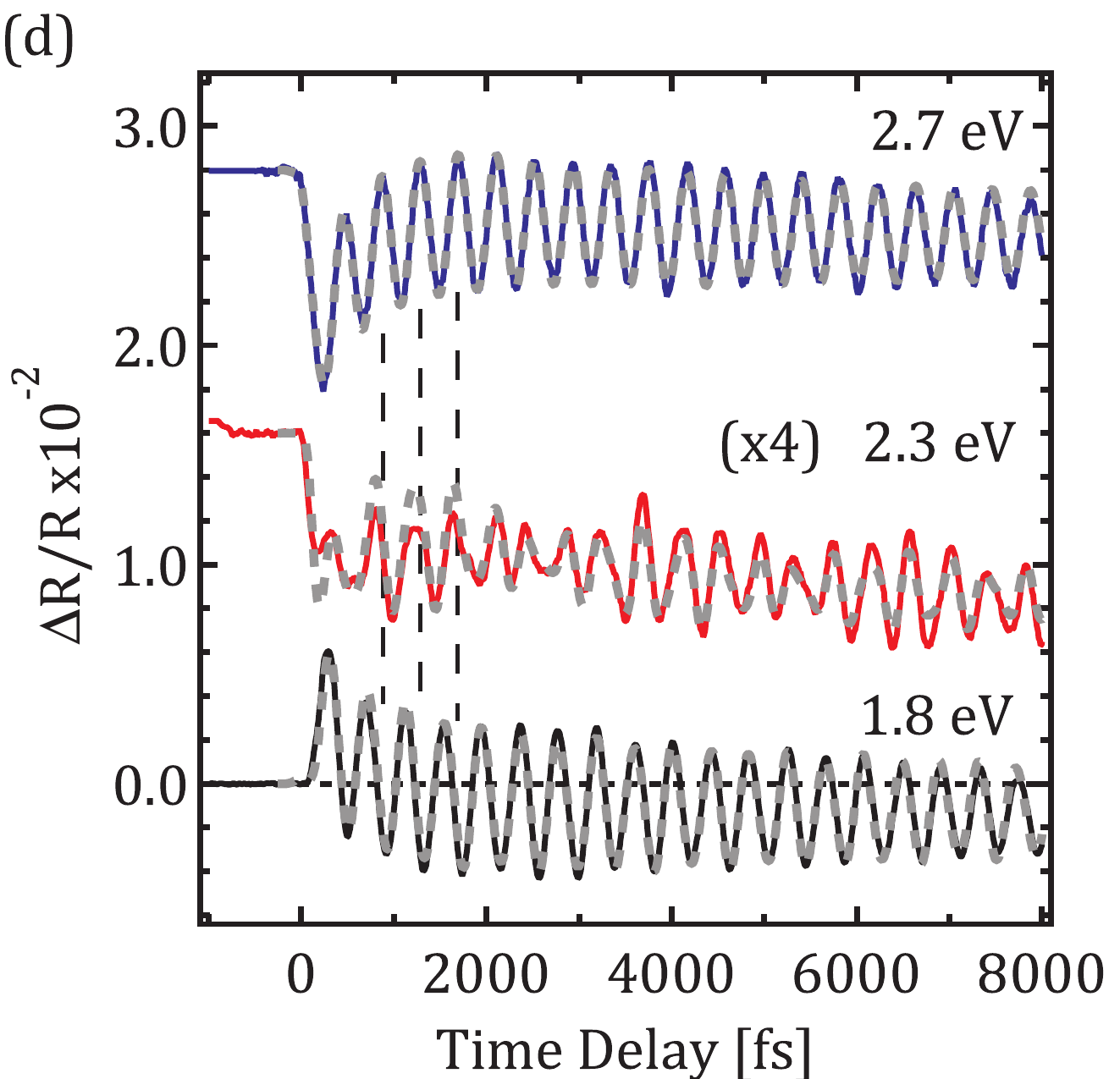}\hfill
	\includegraphics[width=.25\linewidth]{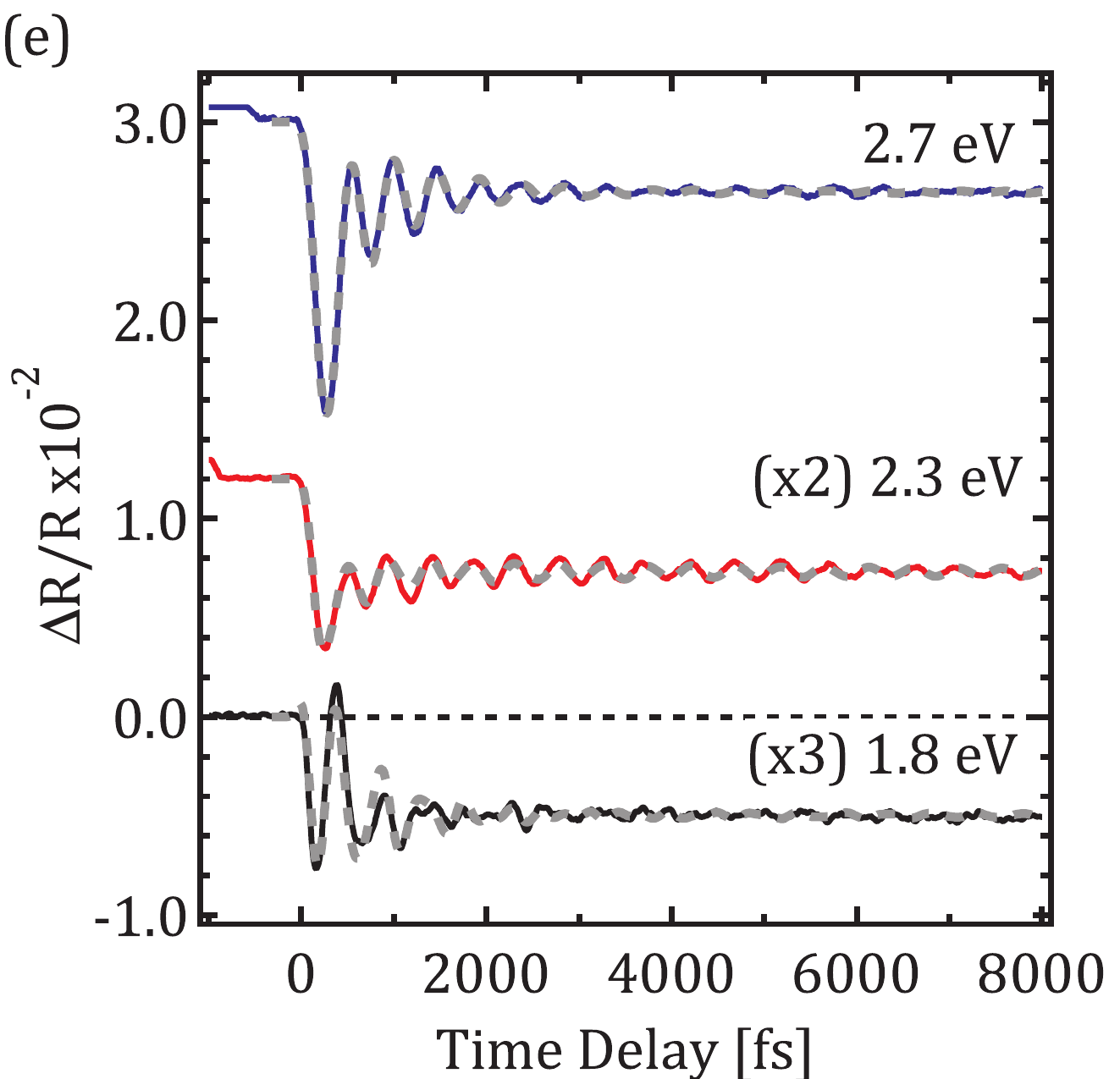}\hfill
	\includegraphics[width=.25\linewidth]{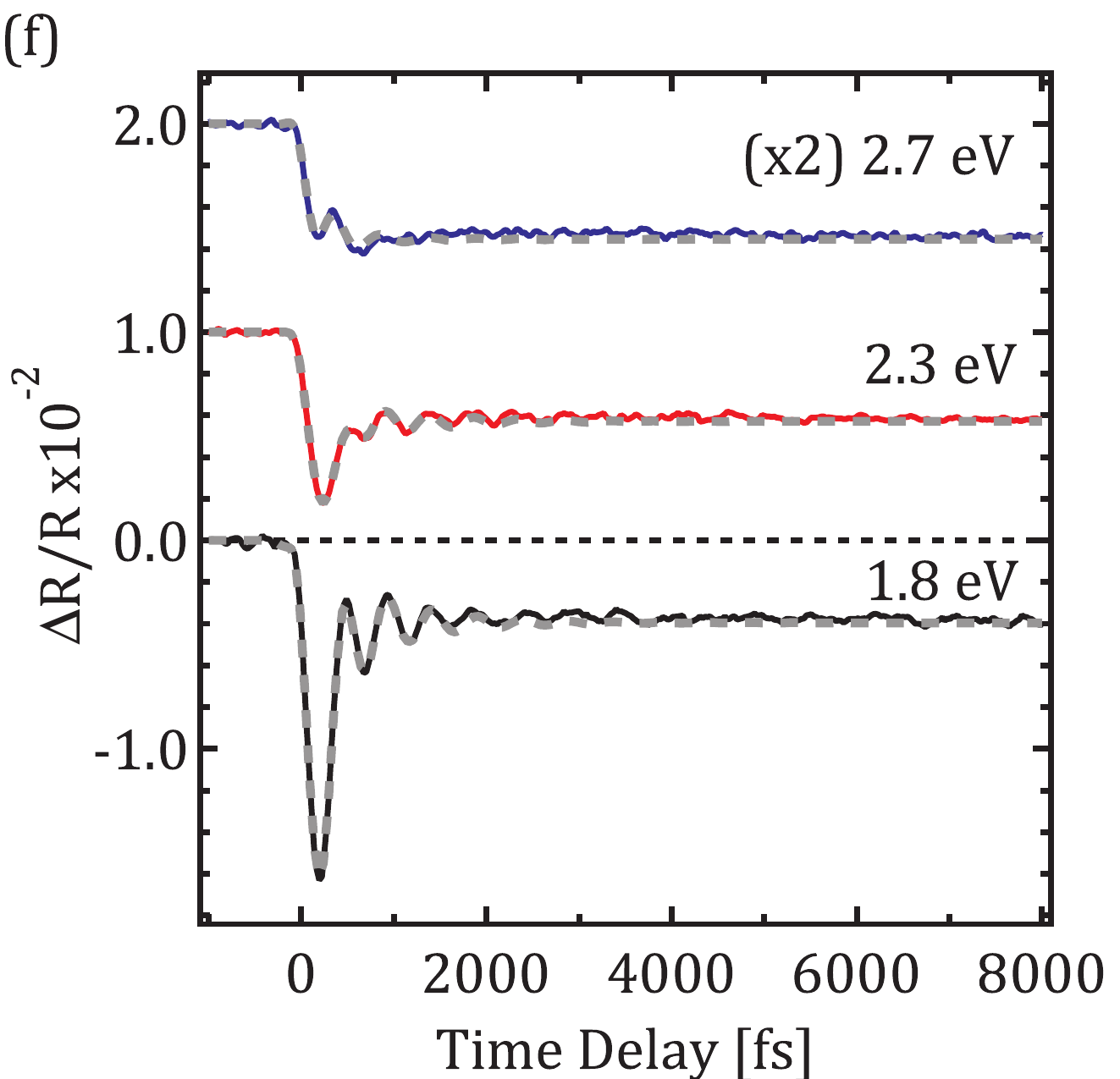}\hfill
	\includegraphics[width=.25\linewidth]{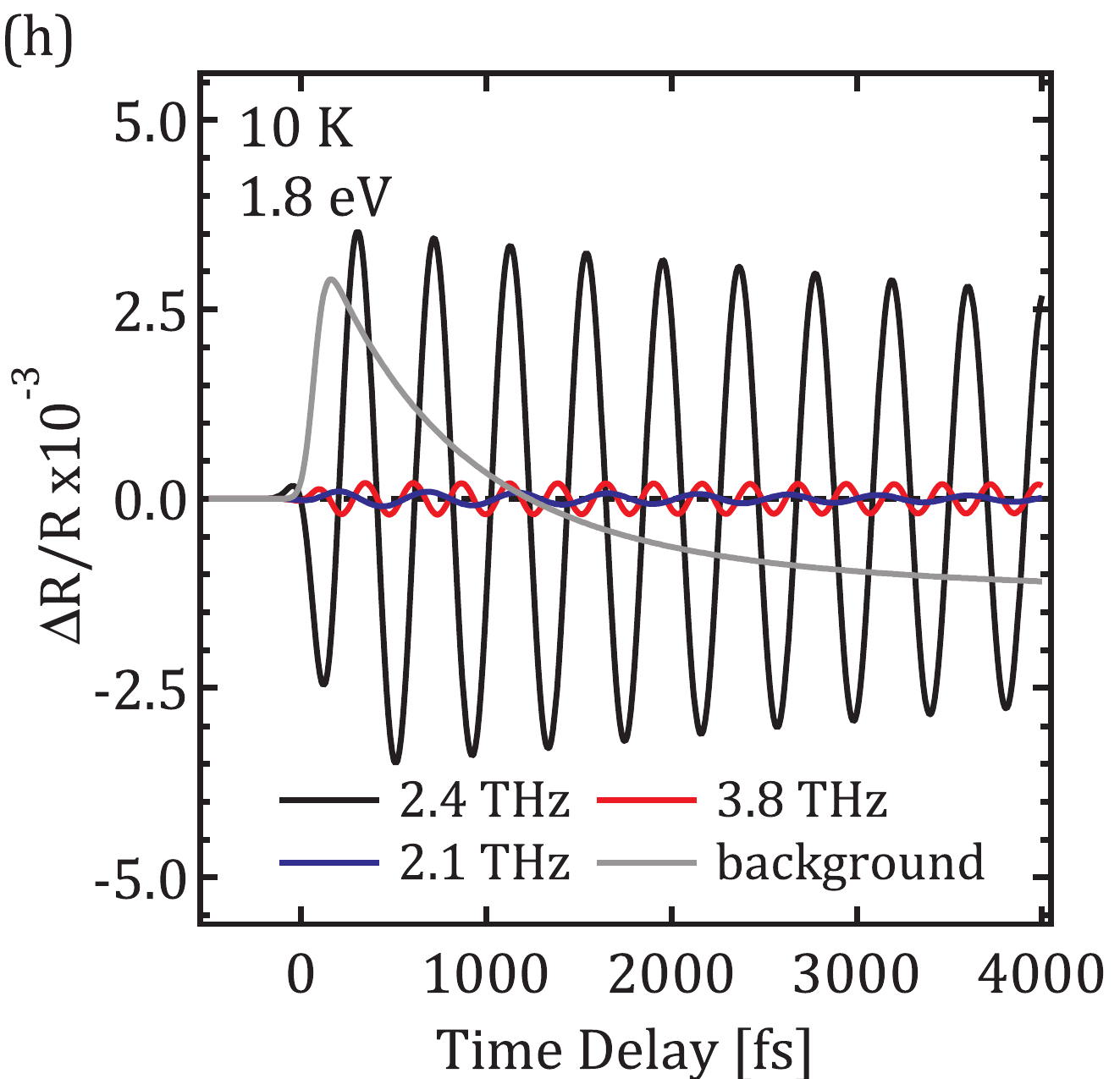}
	\caption[Time-resolved reflectivity spectra of TaS2 in the c-CDW and n-CDW phases]{
	(a)-(c): Time-resolved reflectivity spectra of TaS$_2$ in the c-CDW (10~K) and n-CDW (230~K, 300~K) phases. 
	(d)-(f): Time traces for probe energies of 1.8~eV, 2.3~eV and 2.7~eV. 
	The results of the global fit analysis are overlaid as broken lines. 
	The vertical lines in panel (d) indicate the phase change of the amplitude mode. 
	Traces have been offset vertically for clarity, and some traces have been scaled. 
	(g) Fourier transform of the traces at 10~K from Fig.~\ref{fig:Fig1}(d). 
	(h)  Decomposition of the fit trace for 1.8~eV probe energy from panel (d). 
	}
	\label{fig:Fig1}
\end{figure*}

\section{Methods}
The TaS$_2$ samples were grown by \emph{chemical vapor deposition}~\cite{Dardel1992}: 
Single crystalline samples were prepared from pure elements with iodine as a transport agent, with the growth temperature varying between 950$^\circ$C in the hot zone and 900$^\circ$C in the cold zone. 
The 1$T$-polytype was stabilized by the addition of SnS$_2$ (less than 0.5\% weight) and by rapid cooling from the growth temperature. 
The material grows in flakes with a diameter of few mm and a thickness of some 100~$\mu$m. 
The surface of the flakes is perpendicular to the $c$-axis of the material. 
For the time-resolved experiments, flakes of appropriate dimensions were chosen and oriented by Laue diffraction. 
A clean surface was produced by scotch-taping and is parallel to the Ta layers. 

Time-resolved reflectivity measurements were performed under various experimental conditions using a broadband pump-probe setup~\cite{Baldini2016a,Mann2015}: 
The output of a Ti:sapphire amplified laser system providing sub-50~fs pulses at 1.55~eV was split into pump and probe beams. 
The pump beam was used directly for excitation at 1.55~eV, whereas for tunable excitation in the near-IR it was converted using an optical parametric amplifier system capable of providing sub-100~fs pulses with central energies between 0.5~eV and 1.1~eV. 
To generate the broadband probe beam, pulses of 1~$\mu$J energy were focused into a CaF$_2$ crystal, generating a continuum between 1.75 and 2.75~eV. 
The broadband probe beam was focused onto the sample under an angle of about 15$^\circ$ from the surface normal using parabolic mirrors. 
Its reflection was collected via optical fiber into a $f/4$-spectrometer and read out by a linear \textit{complementary metal–-oxide–-semiconductor} (CMOS) detector synchronized to the laser system. 
The sample was kept in a closed-cycle cryostat at a vacuum pressure lower than $10^{-7}$~mbar.

\section{Results}

\subsection{Temperature dependence}

\begin{figure*}[t]
	\includegraphics[width=.25\linewidth]{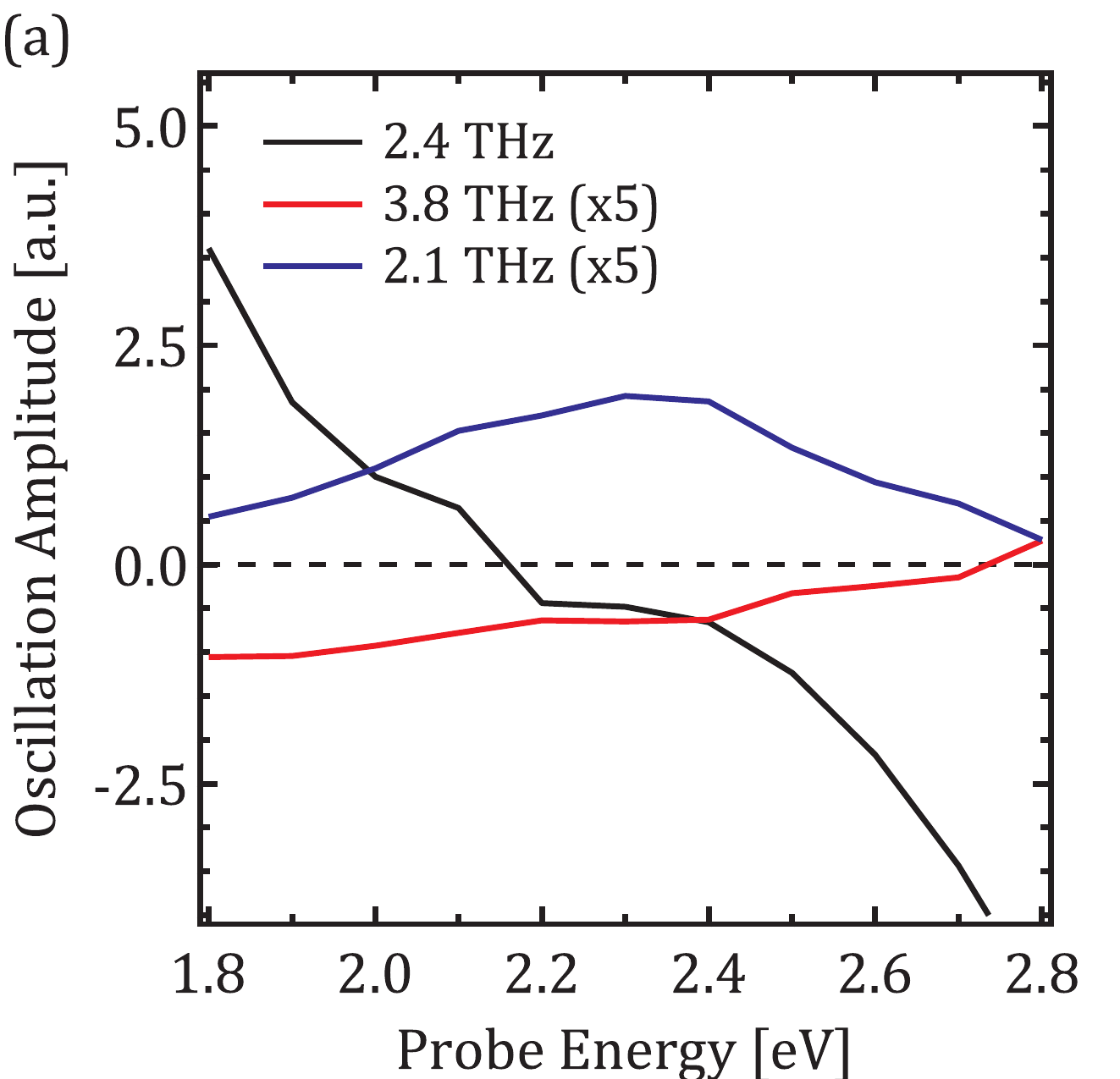}\hfill
	\includegraphics[width=.25\linewidth]{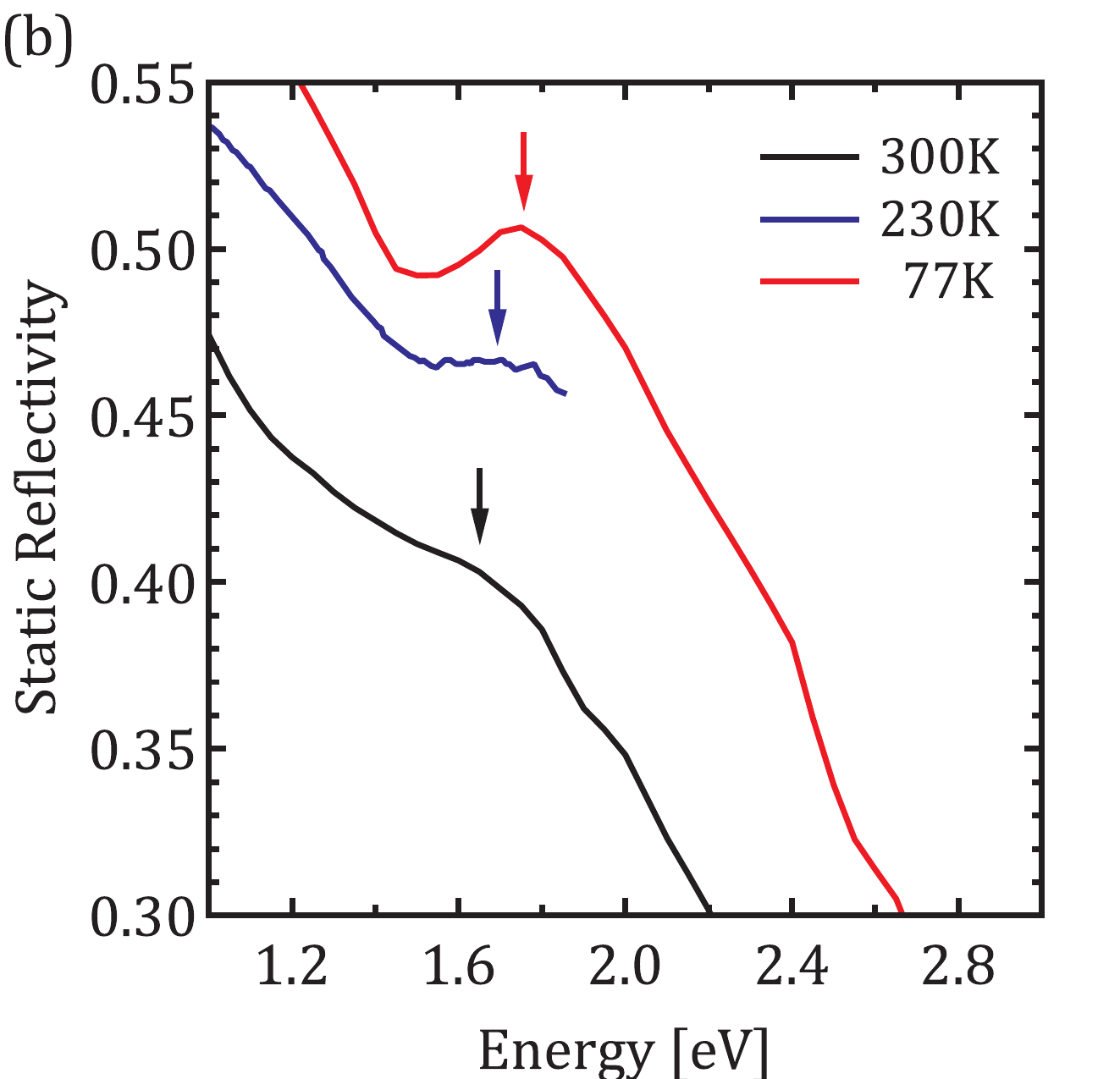}\hfill
	\includegraphics[width=.25\linewidth]{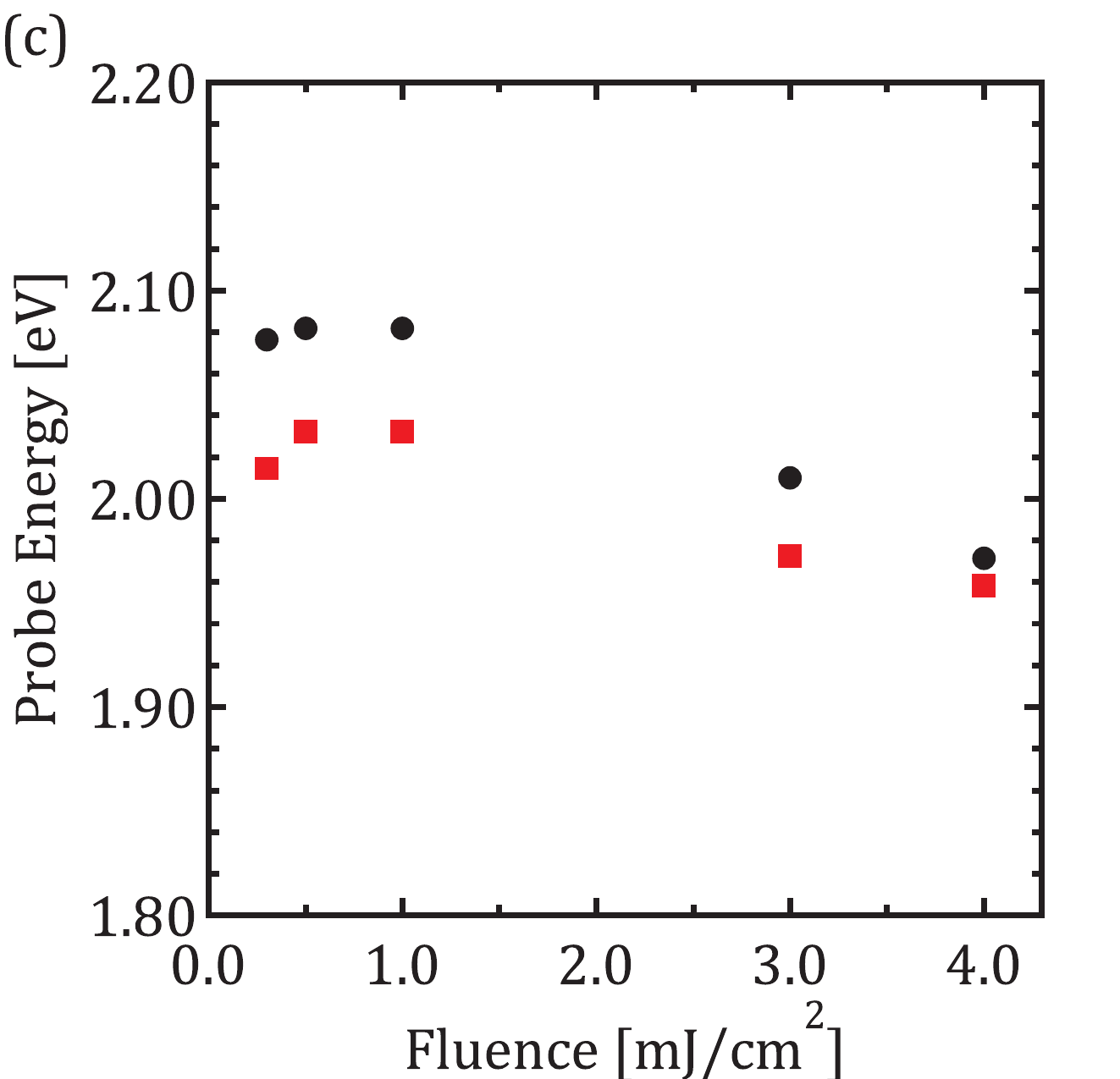}\hfill
	\includegraphics[width=.25\linewidth]{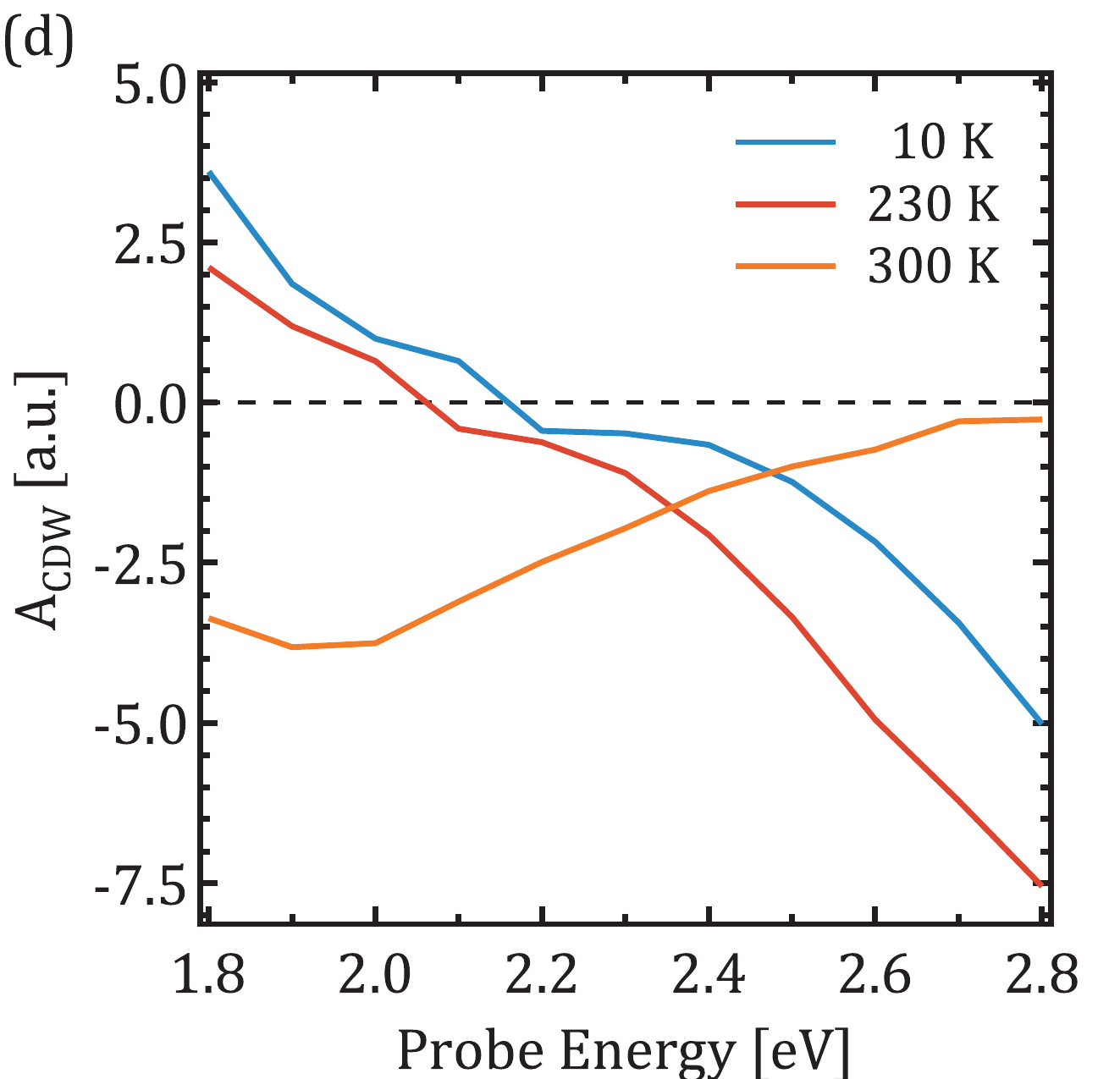}
	\caption[Spectral dependence of the coherent phonon amplitudes in TaS2]{
	(a) Spectral dependence of the coherent phonon amplitudes for all three excited modes in the c-CDW phase at 10~K. 
	(b) Detail of the static reflectivity of TaS$_2$  at 77~K and 300~K~\cite{Beal1975} and at 230~K~\cite{Dean2011}. 
	The arrows indicate the redshift of the broad peak around 2~eV. 
	(c) Fluence dependence of the zero-crossing of the amplitude mode (circles) and the short-lived component of the incoherent background (circles) at 10~K. 
	(d) Temperature dependence of the CDW amplitude mode, i.e. the 2.4~THz mode at 10~K and the 2.1~THz mode at 230~K and 300~K. 
	}
	\label{fig:Fig2}
\end{figure*}

The time-resolved reflectivity change of the sample for an excitation energy of 1.55~eV at the three different temperatures is shown in Fig.~\ref{fig:Fig1}. 
The absorbed pump fluence is 1~mJ/cm$^2$. 
Panels (a)--(c) are color-coded maps showing the reflectivity change as a function of time delay and probe photon energy in the c-CDW phase (10~K) and the n-CDW phase (230~K, 300~K). 
Panels (d)--(f) show the corresponding reflectivity change as a function of time at fixed probe energies (1.8~eV, 2.3~eV, 2.7~eV). 
The overall signal is composed of strong oscillations (visible as vertical lines in the maps) overlaying an incoherent background. 
The oscillations correspond to three different coherent phonons related to the CDW superlattice. 
The incoherent background comprises a short-lived electronic contribution that decays on the el-ph coupling time scale ($\sim 0.5$~ps)~\cite{Stojchevska2014} and a long-lived ($\gg 10$~ps), slow-rising contribution that is of thermal origin. 
We point out that static reflectivity measurements (cf. Fig.~\ref{fig:Fig2}(b)) show a clear absorption feature within our probe spectrum, around 2~eV. 
While its microscopic origin is not assigned in the literature, a close inspection of the band structure~\cite{Smith1985,Rossnagel2006} suggests that this peak is representative of a manifold of Ta $5d$ interband transitions, making our measurements sensitive to the ionic displacements associated with the CDW and its related phonons. 

In accordance with earlier reports~\cite{Demsar2002}, we find no change of the oscillations upon changing the pump and probe polarizations. 
The amplitude mode oscillation has a cosine shape with respect to the exponential background which agrees with an excitation via displacive excitation of coherent phonons (DECP) as suggested in previous reports~\cite{Dean2011,Laulhe2015}. 

The different components of the transient reflectivity signal are most clearly distinguished in the data taken at 10~K, where the coherence time of the phonons is far greater than the probed temporal window of 8~ps. 
A Fourier transform of the time traces at this temperature, shown in Fig.~\ref{fig:Fig1}(h), reveals the presence of coherent modes at 2.4~THz and 3.8~THz, corresponding to the CDW amplitude mode and an additional totally symmetric phonon mode, respectively~\cite{Sugai1981,Demsar2002,Toda2004}. 
The slow-rising 2.1~THz mode that has been observed in single-wavelength pump-probe experiments~\cite{Toda2004} is present in the data as well, but is merged in the Fourier transform with the amplitude mode peak. 
Its presence is the cause of the beating that is clearly visible in the data at 2.3~eV probe energy in Fig.~\ref{fig:Fig1}(d). 
Apart from the approximate phonon frequencies, the Fourier transform offers little quantitative information on the actual mode amplitudes and decay times. 
This information was instead obtained using a global fit procedure: 
The spectra were divided into twelve time traces by integrating over energy windows of 100~meV width. 
A single fit function was used to fit \emph{all} measured reflectivity spectra over their whole probe spectrum with very good accuracy. 
It comprises exponentials modeling the short-lived and long-lived components of the incoherent background, and three damped oscillations to account for the observed phonons. 
For each dataset, the global fit function uses shared rise and decay times and oscillation frequencies across the probe spectrum. 
The fit results are overlaid as broken lines to the time traces in Fig.~\ref{fig:Fig1}(d--f). 
Figure~\ref{fig:Fig1}(h) illustrates the different components of the model function. 
The main information obtained from the global fit is the spectral dependence of the different fit components, which carries information on their resonant behavior. 

In the following, we discuss the details of the coherent oscillations and the incoherent background in the c-CDW phase at 10~K, before commenting on the changes in the n-CDW phase. 
The probe energy dependence of the excited phonons at 10~K that is obtained from the global fit of the data in Fig.~\ref{fig:Fig1}(a) and (d) is shown in Fig.~\ref{fig:Fig2}(a) in terms of the oscillation amplitudes: 
The dominant CDW amplitude mode at 2.4~THz amplitude mode exhibits a phase change of $\pi$, indicated by the vertical broken lines in Fig.~\ref{fig:Fig1}(d), that occurs around 2.1~eV. 
This is consistent with a deformation potential effect~\cite{Bardeen1950}, in which the periodic lattice distortion modulates the central energy of an absorption peak around 2~eV which is observed in the static reflectivity (Fig.~\ref{fig:Fig2}(b)). 
The shift of the peak's central energy leads to a differential reflectivity with a phase shift close to the peak position. 
A slight deviation from the exact energy of the static peak can be attributed to a slightly modified electronic structure of our sample with respect to the literature data, or to a broadening effect. 
The amplitude displays a kink close to the zero-crossing, which is an effect of the comparatively wide integration window (100~meV) that is used in the global fit, leading to a data point that integrates over the region where the amplitude crosses zero and the phase of the phonon is not well defined. 
The initial peak of the amplitude mode oscillation is negative, indicating that the probed absorption peak redshifts upon photoexcitation, as is expected. 
Furthermore, the intensity of the CDW amplitude increases sharply on both sides of the probe spectrum. 
The modes 2.1~THz and 3.8~THz have a rather small amplitude, but a very long coherence time, causing them to be visible in the Fourier transform as well as the global fit analysis. 
The energy dependence of the phonon amplitudes explains why the CDW amplitude mode is highly dominant in single-wavelength time-resolved experiments at 1.55~eV~\cite{Demsar2002,Onozaki2007,Eichberger2010}. 
The 2.1~THz mode shows a peak in the center of the spectrum; close to the zero-crossing of the amplitude mode, in accordance with the time traces in Fig.~\ref{fig:Fig1}(d). 
The mode is thus readily observable in the center of our probed spectral range even for short delay times. 
In contrast, in single-wavelength experiments at 1.55~eV the  relative strength of the mode is low for the first few ps~\cite{Toda2004}. 
There, it was speculated that the mode was not observed at short time delays because of the nature of its excitation mechanism. 
The broadband nature of our probe allows us to clarify this point, as the mode is visible at early time delays for probe energies around 2.3~eV. 
We point out that the shape of the 2.1~THz mode suggests a different generation mechanism as for the amplitude mode: 
The peaked shape indicates an amplitude modulation of the absorption peak, which is caused by the photoinduced change in the density of states, or a change of the transition dipole matrix element. 
Therefore, one would expect the zero-crossing of the amplitude mode to coincide with the peak of the 2.1~THz mode, while Fig.~\ref{fig:Fig2}(a) indicates a slight deviation. 
This is most likely a nonlinear effect. 
The 3.8~THz mode displays a flat spectral response at a low amplitude, and we can draw no conclusion as to its origin. 

We continue with the incoherent background at 10~K. 
The long-lived component of the background exhibits a flat, negative amplitude. 
Seeing that the static reflectivity in Fig.~\ref{fig:Fig2}(b) decreases across our probe spectrum with increasing temperature, we identify the long-lived incoherent background as a thermal contribution. 
Interestingly, the short-lived component of the background behaves in a fashion that is very similar to the amplitude mode, inverting its sign over the probe spectrum, as seen in Fig.~\ref{fig:Fig1}(a) and (d). 
This is consistent with a carrier density-dependent renormalization of the underlying interband transition. 
In accordance with this interpretation, the zero-crossing of the electronic background and the CDW mode amplitude coincide. 
We determine the position of the zero-crossing as the minimum of the absolute value of the amplitude in order to increase the fit accuracy and avoid the problem of the kink exhibited close to the amplitude mode zero-crossing. 
Figure~\ref{fig:Fig2}(c) shows the estimated zero-crossing of the electronic background and the amplitude mode at 10~K for various pump fluences. 
Both curves are very similar, although the exact value differs by about two percent. 
This difference is most likely due to the very strong variation of the electronic temperature during the el-ph equilibration, which slightly deforms the spectral shape of the incoherent background as a function of time for the first few hundred fs. 

\begin{figure}[b]
	\includegraphics[width=.33\linewidth]{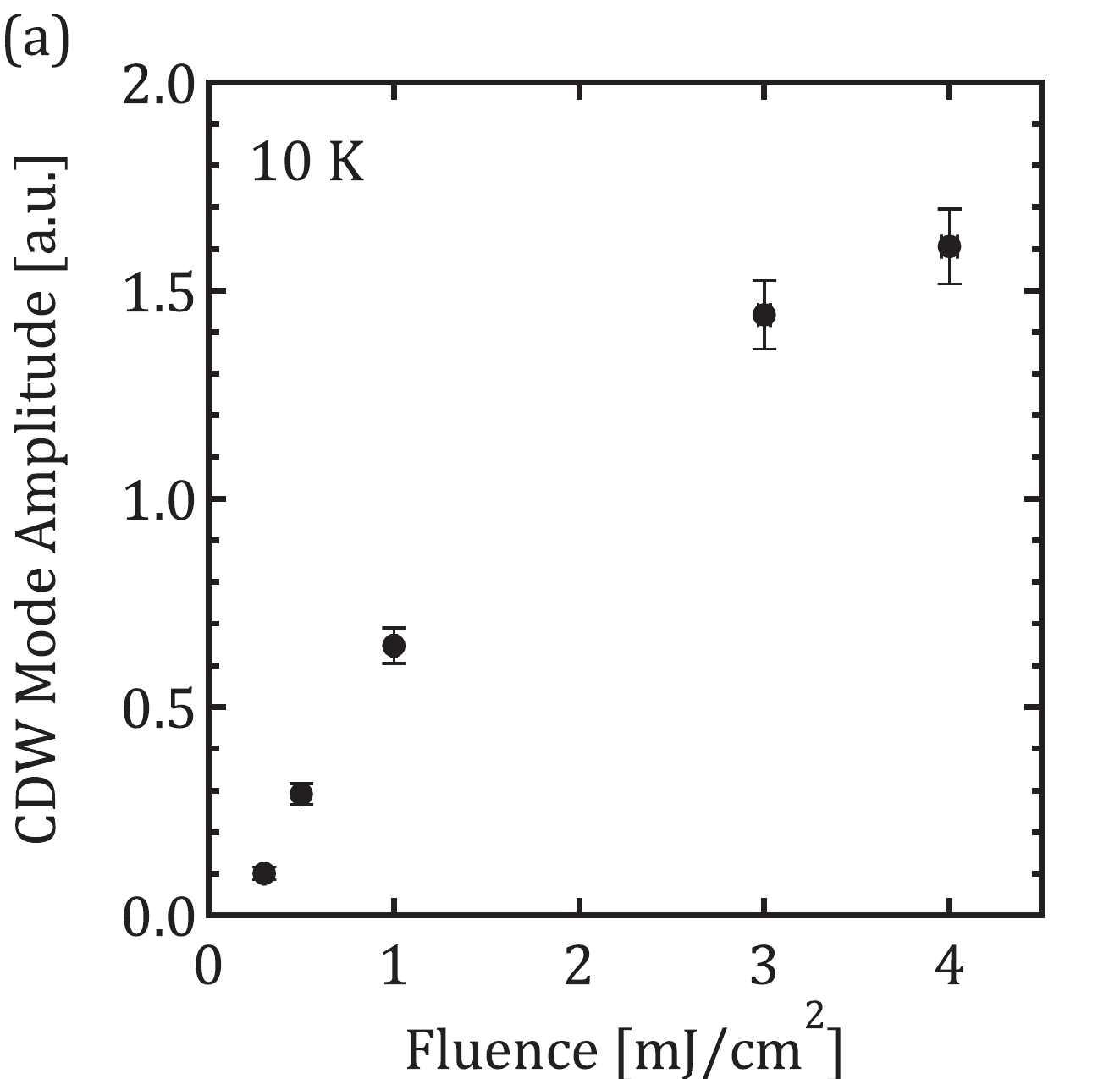}
	\includegraphics[width=.33\linewidth]{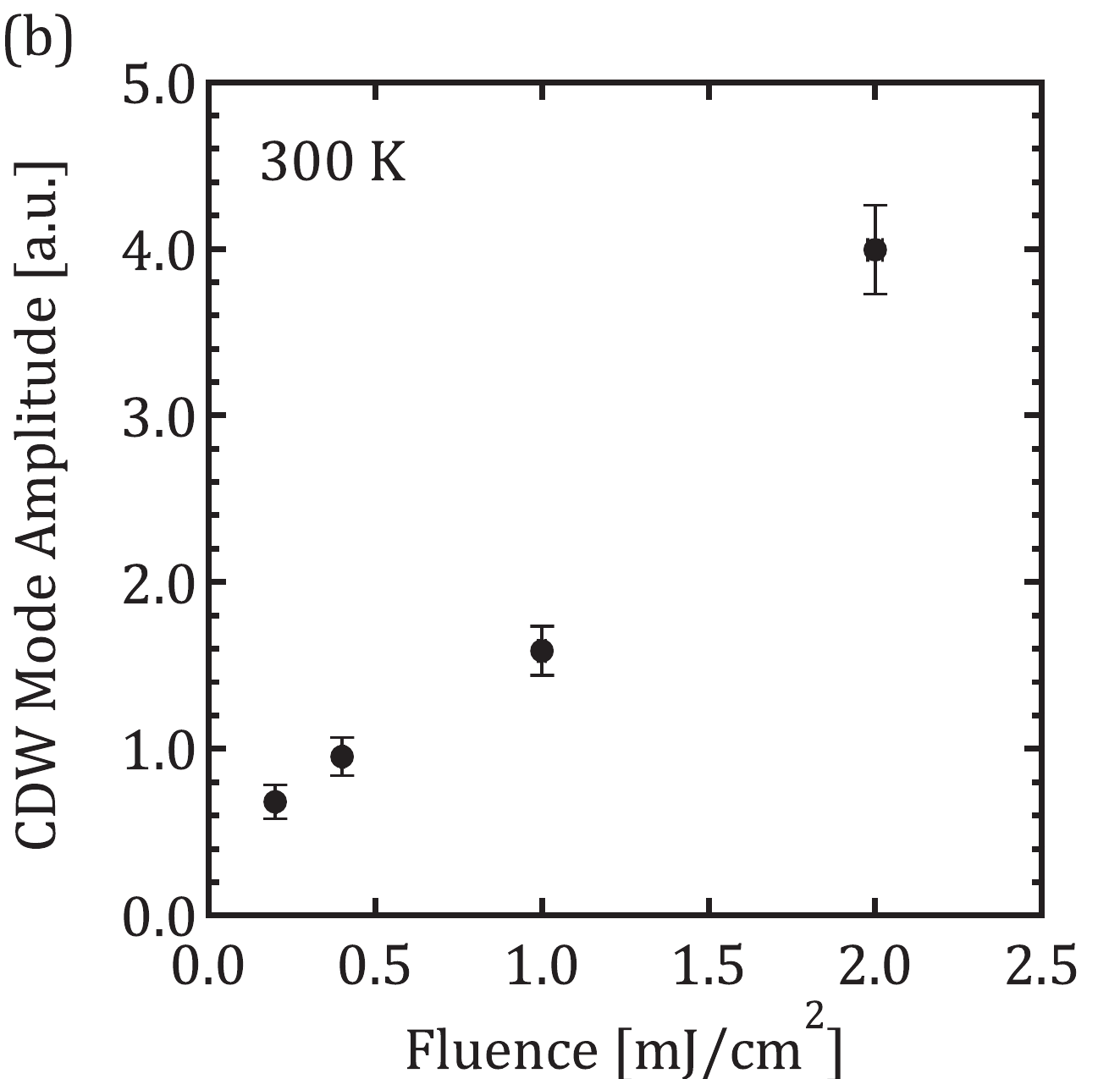}\hspace{0pt}\hfill
	\caption[Fluence dependence of the CDW mode amplitude]{
	Fluence dependence of the CDW mode amplitude close to its maximum value, i.e. at (a) 10~K at 2.7~eV probe energy and (b) at 300~K at 2.0~eV probe energy, as extracted from the global fit. 
	}
	\label{fig:Fig3}
\end{figure}
\begin{figure*}[tb]
	\includegraphics[width=0.33\linewidth]{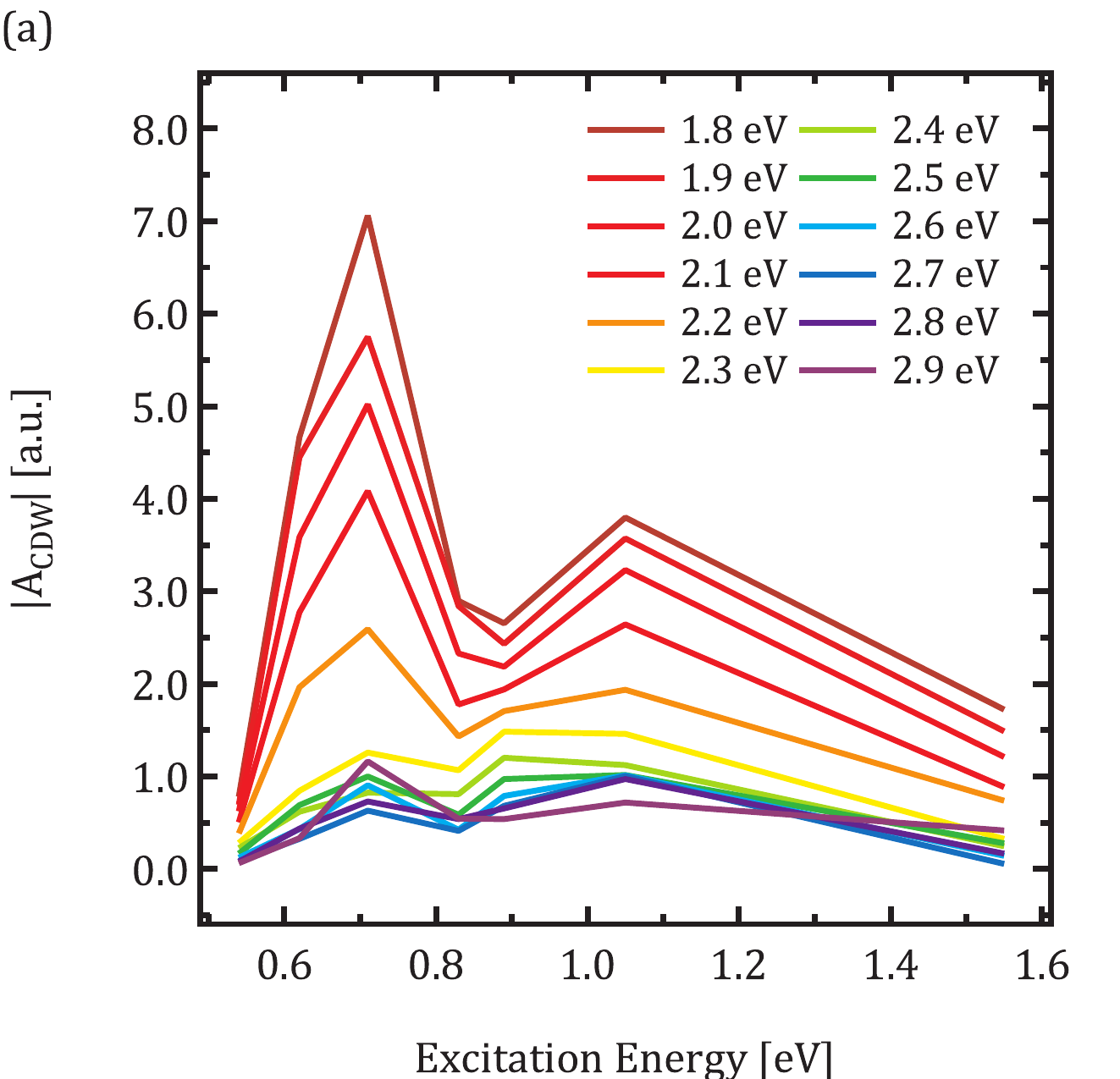}\hfill
	\includegraphics[width=0.33\linewidth]{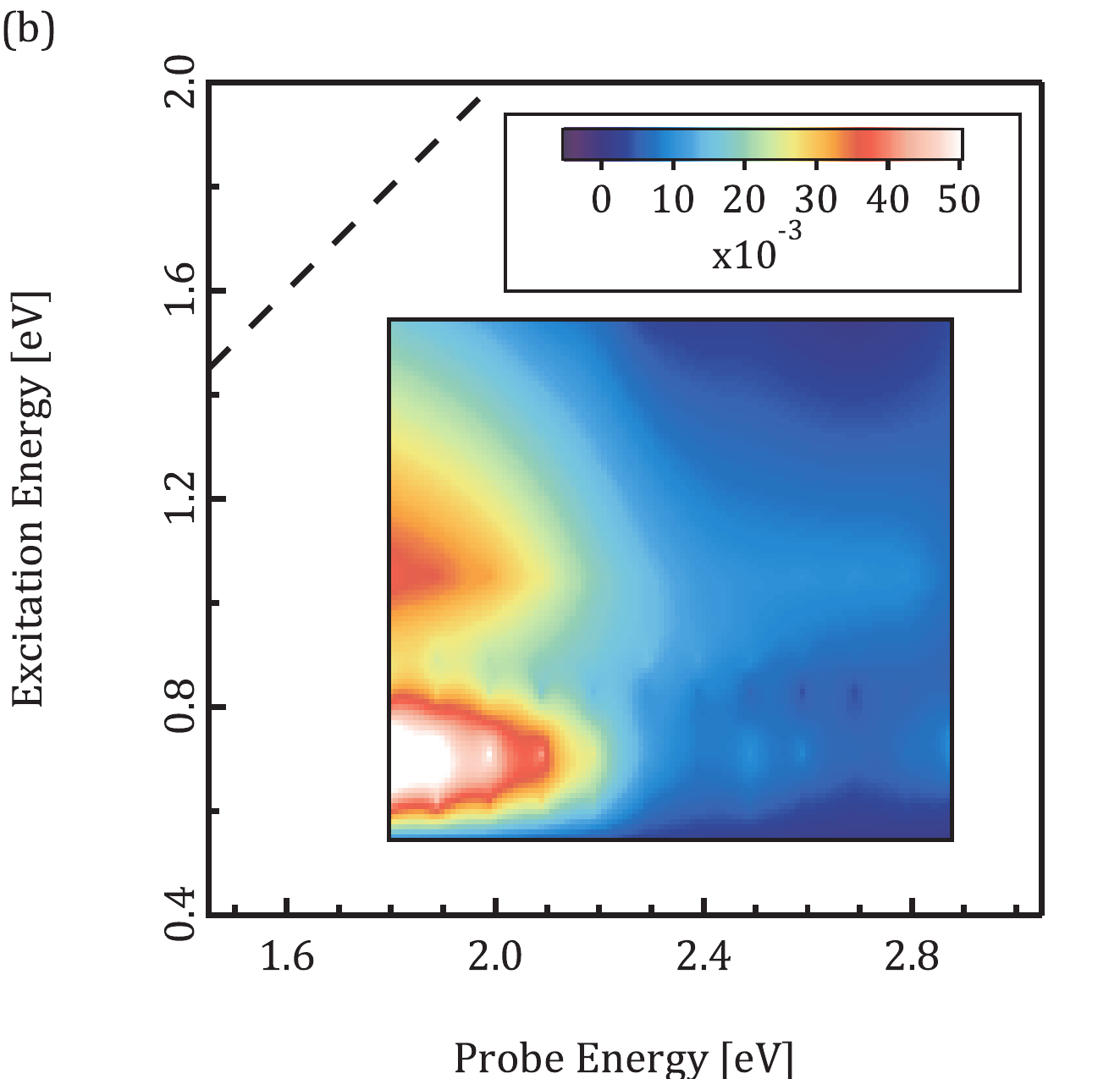}\hfill
	\includegraphics[width=0.33\linewidth]{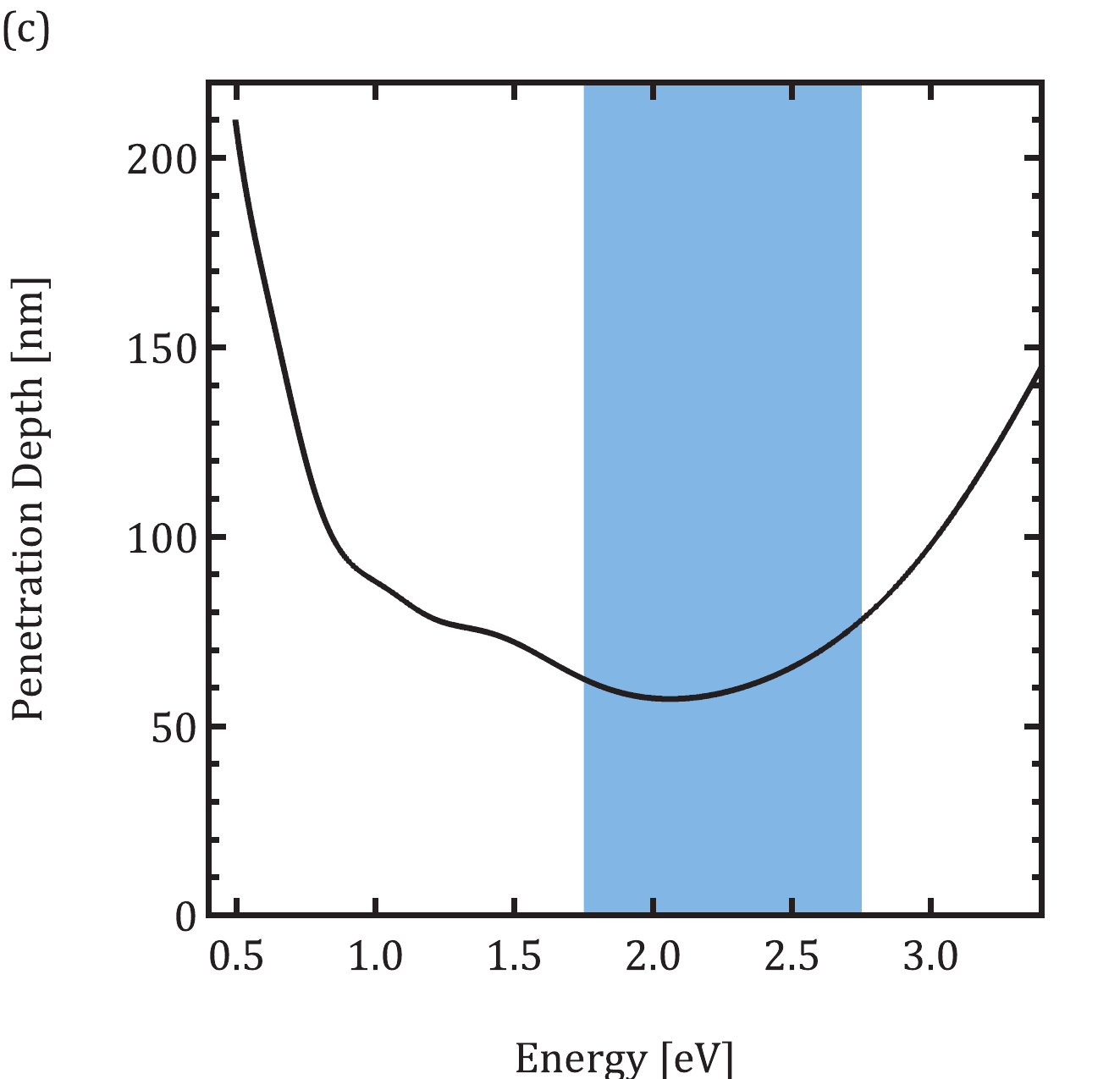}
	\caption[Resonant behavior of the CDW amplitude as a function of the excitation energy in the IR]{
	(a) Resonant behavior of the CDW amplitude as a function of the excitation energy in the IR at 300~K, for varying probe energies. 
	All curves have been normalized according to the absorbed fluence and penetration depth mismatch. 
	For simplicity, absolute values are shown. 
	(b) Two-dimensional image of the data from (a), with the amplitude given in color-coding, showing the resonances for excitation energies of 0.7\,eV and 1.1\,eV. 
	(c) Penetration depth $\delta$ of TaS$_2$, calculated from static conductivity data~\cite{Dean2011}. 
	The shaded area denotes our probe spectrum. 
	}
	\label{fig:Fig4}
\end{figure*}
We now address the analysis of the data in the n-CDW phase at 230~K from Fig.~\ref{fig:Fig1}(b)--(c) and (e)--(f). 
The transition to the n-CDW phase is marked by a significant reduction in the coherence time of the phonons, with a strong damping during the first ps. 
The CDW amplitude mode decreases its frequency to 2.1~THz, while the mode at 3.8~THz softens to 3.6~THz, in accordance with spontaneous Raman data~\cite{Sugai1981,Sai-Halasz1977}. 
As in the Raman data, the 2.1~THz from the c-CDW phase can no longer be observed independently, and is most likely not present. 
The incoherent background at 230~K is similar to the c-CDW phase, but its zero-crossing is slightly redshifted. 
At 300~K, it is instead strongly peaked around 2~eV, and purely negative. 
Both observations are consistent with the static reflectivity from Fig.~\ref{fig:Fig2}(b), where the observed reflectivity feature is redshifted at 230~K with respect to the c-CDW phase, while it at 300~K it is broadened the point where no peak (and thus, no zero-crossing in the differential reflectivity) is seen. 
The long-lived component of the background is negative over the whole probe range in the n-CDW phase, and slightly gains in amplitude. 
This again is consistent with the trends observed in the static reflectivity. 
The temperature dependence of the amplitude mode is shown in Fig.~\ref{fig:Fig2}(d). 
The dependence is nearly identical for 10~K and 230~K  apart from the aforementioned redshift of the zero-crossing, and follows the trend of the electronic background. 
Likewise, the profile at 300~K resembles the shape of the incoherent peak visible in the spectrum in Fig.~\ref{fig:Fig1}(c). 

It should be noted that at 10~K the sample response is linear in pump fluence for absorbed fluences smaller than 1~mJ/cm$^2$, while showing saturation for greater fluences. 
This is shown for the CDW amplitude mode in Fig.~\ref{fig:Fig3}(a). 
At 300~K (Fig.~\ref{fig:Fig3}(b)) the fluence dependence is more complex: 
A saturation is seen up to 1~mJ/cm$^2$. 
At higher fluences the deposited energy is enough to increase the sample temperature above the phase transition to the i-CDW phase. 
An estimate of the maximum sample temperature is given in the appendix. 
The damage threshold of the sample at this temperature is close to a fluence of 4~mJ/cm$^2$. 

Finally, we comment on the value of the pump fluence used in the dataset of Fig.~\ref{fig:Fig1}. 
It has been demonstrated in literature~\cite{Stojchevska2014} that at 10~K a slightly shorter pulse ($\sim 35$~fs) of similar fluence is enough to trigger the transition from the c-CDW phase to a hidden (H) state. 
This state is known to persist much longer than the pulse separation of our setup, raising the question whether or not our pump pulse continuously switches the sample between the two states. 
To this end, we want to point out that the fluence dependence at 10~K does not show a marked discontinuity between 0.3 and 1.0~mJ/cm$^2$, where the threshold is expected to be~\footnote{The full data sets are provided as supplementary material}. 
However, our probe range is not sensitive to the established hallmarks of the photoinduced transition, namely the change in the static reflectivity and the electrical resistivity. 
The slight shift of the amplitude mode frequency from 2.46~THz to 2.39~THz upon the transition to the H state is below our resolution; the frequency of the amplitude mode obtained from our global fit is 2.43~THz. 
It would be interesting to examine to what extent the observed resonant behavior changes upon the transition to the H state, and whether the material shows evidence of the transition in our probe spectrum. 
While it is difficult to reduce the pump fluence below 0.3~mJ/cm$^2$ given our current signal-to-noise ratio, it may be possible to observe an effect when performing pump-probe experiments in a regime where the H state is less stable and the sample definitely recovers between laser pulses.

\subsection{Resonant behavior of the CDW amplitude mode}

In order to establish whether the CDW amplitude mode couples to the high-energy electronic excitations, we tuned the excitation energy in the near-IR range. 
The transient broadband reflectivity data for excitation energies between 0.54~eV and 1.05~eV in the n-CDW phase at 300~K are shown in the appendix. 
All spectra follow the general trend observed for the measurement at 300~K in Fig.~\ref{fig:Fig1}(c): 
The signal peaks around 2~eV probe energy and is dominated by the incoherent background.  
The CDW amplitude mode at 2.1~THz is clearly visible and decays within approximately 1~ps. 

The resonant behavior of the CDW amplitude mode extracted from the broadband data is shown as a function of excitation energy in Fig.~\ref{fig:Fig4}(a). 
All amplitude values in the figure have been rescaled with respect to the absorbed laser fluence and the penetration depth mismatch. 
The penetration depth $\delta$ was calculated from ellipsometry data~\cite{Dean2011}. 
The different curves correspond to different probe energies, and clearly show that the oscillation is strongest when pumping in the near-IR around 0.7~eV and 1.1~eV. 
Figure~\ref{fig:Fig4}(b) displays the resonant behavior of the amplitude mode in the two-dimensional space of excitation and probe energy. 
The amplitude mode exhibits its largest amplitude around 0.7~eV, with a minor, broader peak around 1.1~eV that stretches to higher probe energies. 

Remarkably, the overall resonant behavior seen in Fig.~\ref{fig:Fig4} is devoid of very sharp resonances, in accordance with the displacive excitation mechanism and the fact that all energies used were greater than the Mott gap $\sim~400$~meV, as determined by optical spectroscopy~\cite{Gasparov2002} and numerical methods~\cite{Ritschel2015}. 
Indeed, the optical conductivity of TaS$_2$ exhibits a double peak between 0.5~eV and 1.2~eV~\cite{Gasparov2002}, which supports the interpretation of the data as a resonant behavior of the CDW amplitude mode excited by DECP. 
We note that excitation via the DECP mechanism leads to a resonance profile that should be independent of the probe energy. 
As can be seen in Fig.~\ref{fig:Fig4}(a), this is true for most probe energies, but seems to break down close to the zero-crossing of the amplitude mode. 
This is most likely related to the kink observed in the fits in Fig.~\ref{fig:Fig2} close to the zero-crossing of the phonon amplitudes, or it could indicate that our linear correction is insufficient. 
The peak around 0.7~eV is noteworthy with respect to the recent observation of a long-lived photo-induced conducting state that was excited using 0.8~eV photons~\cite{Laulhe2015}. 

\begin{figure}[tb]
	\hfill
	\includegraphics[scale=0.93]{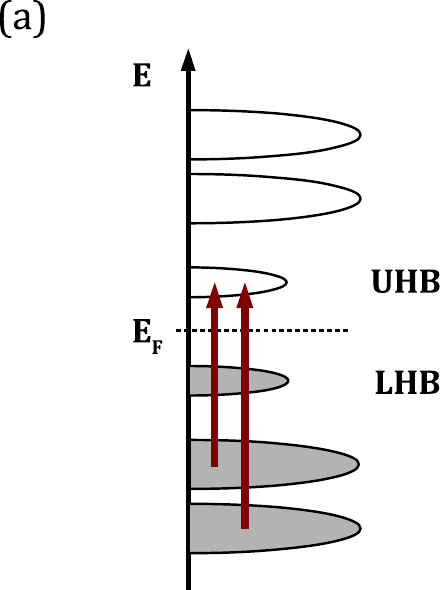}\hfill
	\includegraphics[scale=0.93]{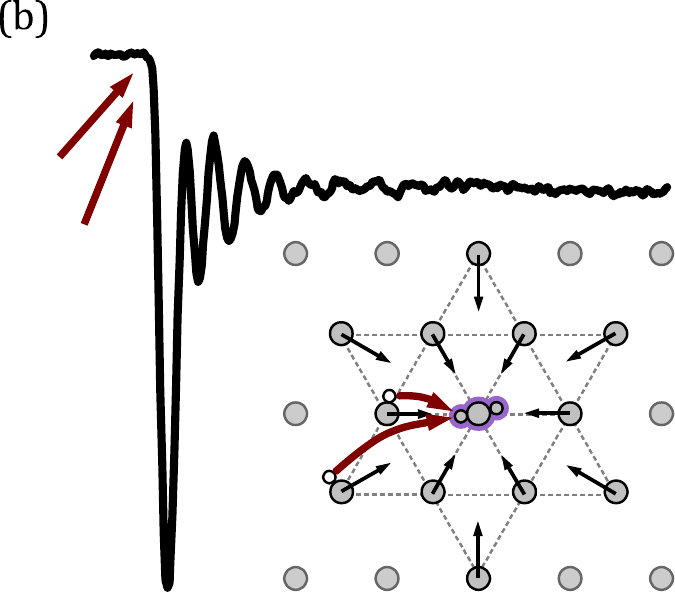}\hfill\hspace{0pt}
	\caption{
	(a) Schematic representation of the density of states around the Fermi energy, based upon ARPES and ARIPES data~\cite{Sato2014}, showing the upper and lower Hubbard band (UHB, LHB) resulting from the Ta atom at the star center, as well as the manifolds associated with states from the twelve outer Ta atoms. 
	Arrows indicate the two transitions associated to the resonance peaks of the CDW amplitude mode excitation. 
	(b) The two transitions correspond to a charge-transfer inside the star-shaped Ta cluster from the outer resp. inner ring to the star center, creating a doublon state that launches the CDW-related phonons. 
	Small arrows indicate the Ta displacement for the CDW amplitude mode. 
	Note that, in reality, the bands are not clearly separated, and hybridized with the orbitals of the sandwiching S atoms, and that the excited electronic states are delocalized. 
	}
	\label{fig:Fig5}
\end{figure}
The strong resonance of the CDW excitation around the two peaks can be explained by examining the band structure of TaS$_2$ in the charge-ordered phase. 
A combination of ARPES~\cite{Arita2004} and ARIPES data~\cite{Sato2014} allows to identify the two peaks at 0.7~eV and 1.1~eV with direct transitions from states originating from the Ta 5$d_{z^2}$ bands that are formed by the outer Ta atoms on the David's star cluster to the upper Hubbard band that is formed by the unoccupied states contributed by the central Ta atoms of the star (Fig.~\ref{fig:Fig5}(a)). 
These excitations both result in the formation of a doublon at the star center, which strongly couples to the CDW-related phonon modes, especially the amplitude mode (Fig.~\ref{fig:Fig5}(b)). 
We point out that the important change in the band structure occurs upon the transition from the i-CDW phase to the n-CDW phase~\cite{Claessen1990,Sato2014}. 
Therefore, we consider the application of arguments based on the c-CDW band structure justified for our measurements at 300~K. 
The observed features are expected to be much more pronounced in the c-CDW phase.

\section{Summary}
In conclusion, we have performed ultrafast broadband spectroscopy on 1$T$-TaS$_2$, tuning the optical excitation in the near-IR while probing in the visible. 
The optical excitation triggers different coherent collective modes related to the CDW present in the material, the strongest being the CDW amplitude mode at 2.4~THz in the c-CDW phase. 
The observed cosine shape of the oscillations and their resonant behavior agree with a DECP generation mechanism, requiring a driving force that lasts significantly longer than the phonon period. 
Using the broadband probe, the spectral dependence of the modes can be studied over a wide energy range using a global fit procedure. 
The global fit analysis of the data allows to identify the resonant behavior of three coherent modes that are excited. 

We find that the CDW amplitude mode strongly resonates with a doublon state that is created at the center of the charge-ordered stars of Ta atoms, at energies around 0.7~eV and 1.1~eV, whereas the corresponding holon at the outer Ta atoms plays only a minor role. 
The observed resonances are sharp enough to be distinctly observed within our experimental energy range. 
The band structure in the charge-ordered phase of TaS$_2$ allows to identify the resonances as direct transitions related to a charge-transfer within the star-shaped charge-ordered clusters, namely from the outer Ta atoms to the central atom, creating a doublon state at the center and a holon on the outer Ta atom. 
Both resonances correspond to excitations reaching the same final state, indicating that the doublon creation is crucial for launching the CDW mode via the DECP mechanism, whereas the holon has far less influence on the oscillation. 
We would like to point out that the lifetime of the doublon excitation exceeds our probe interval of 8~ps, suggesting that its creation plays a role in the stabilization of ``hidden'' metastable phases that were shown to exist in TaS$_2$~\cite{Stojchevska2014,Laulhe2015}.

\begin{acknowledgments}
	Work at LUMES was supported by ERC starting grant USED258697 (F.C.) and the NCCR MUST, a research instrument of the Swiss National Science Foundation (SNSF). 
	We thank M. Ligges for helpful discussions. 
\end{acknowledgments}

\bibliography{TaS2}

\newpage

\section{Appendix}

\subsection{Fourier transforms of data at 1.55~eV excitation energy}
\begin{figure}[t]
	\includegraphics[width=.33\linewidth]{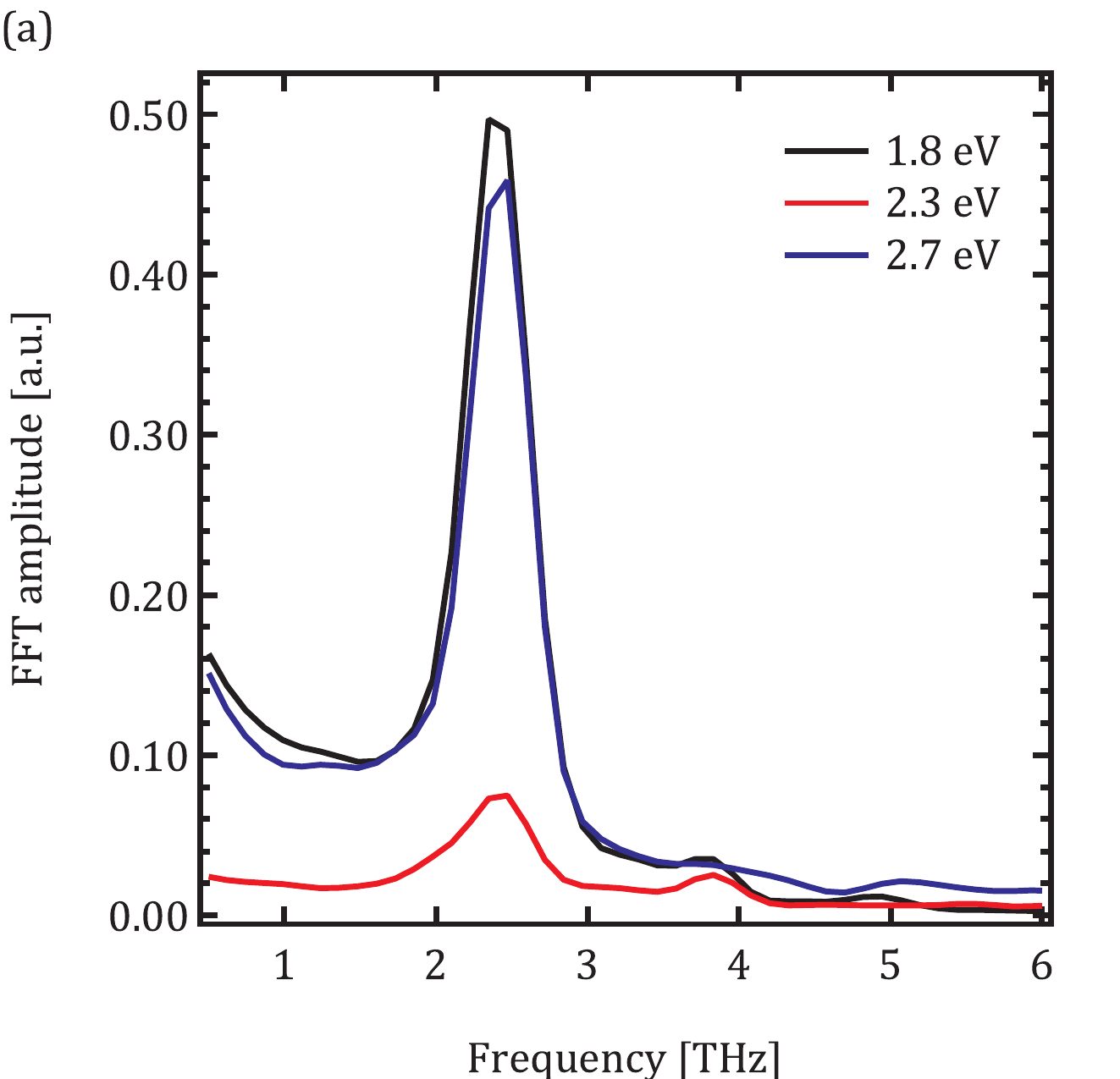}\hfill
	\includegraphics[width=.33\linewidth]{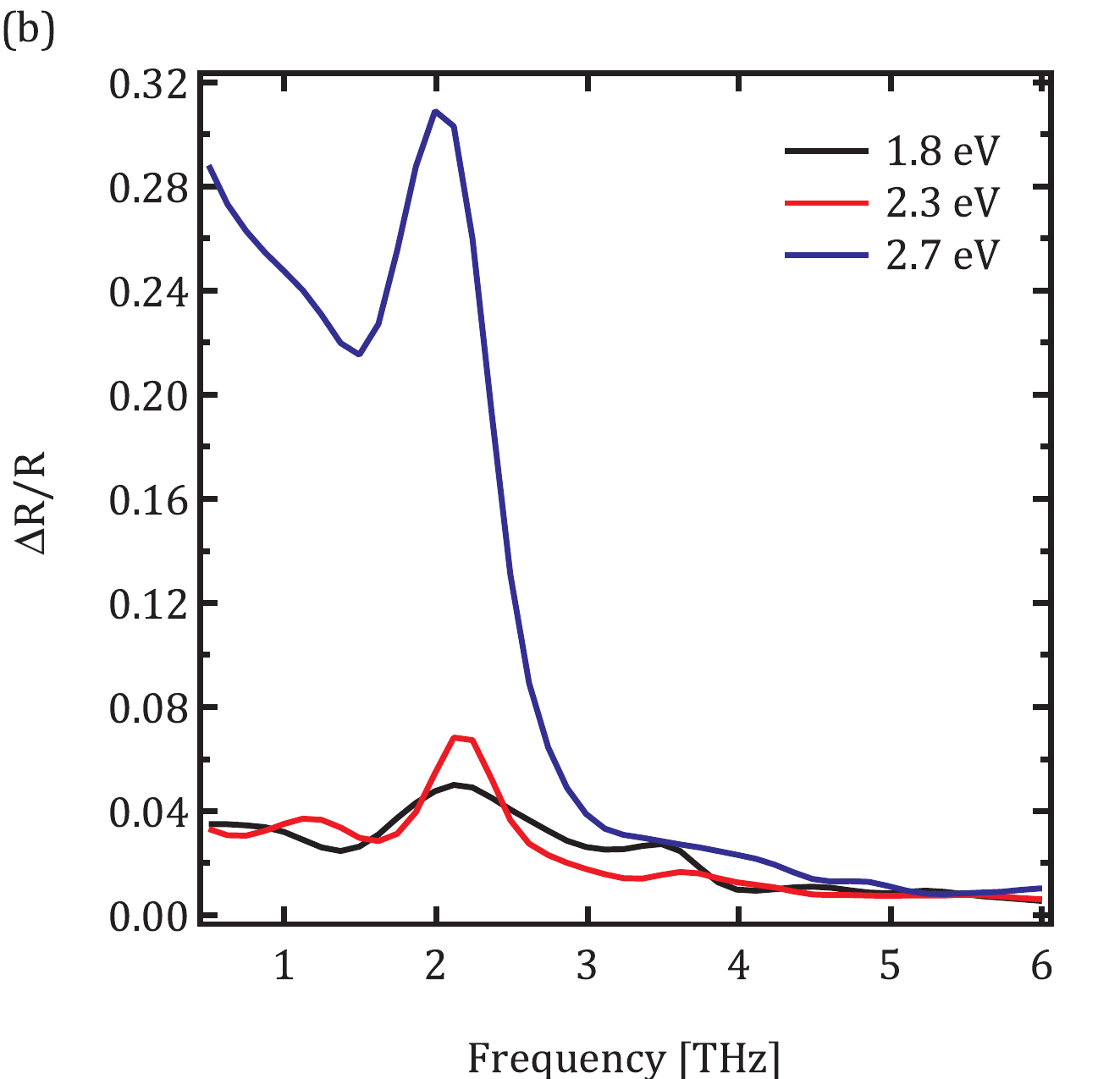}\hfill
	\includegraphics[width=.33\linewidth]{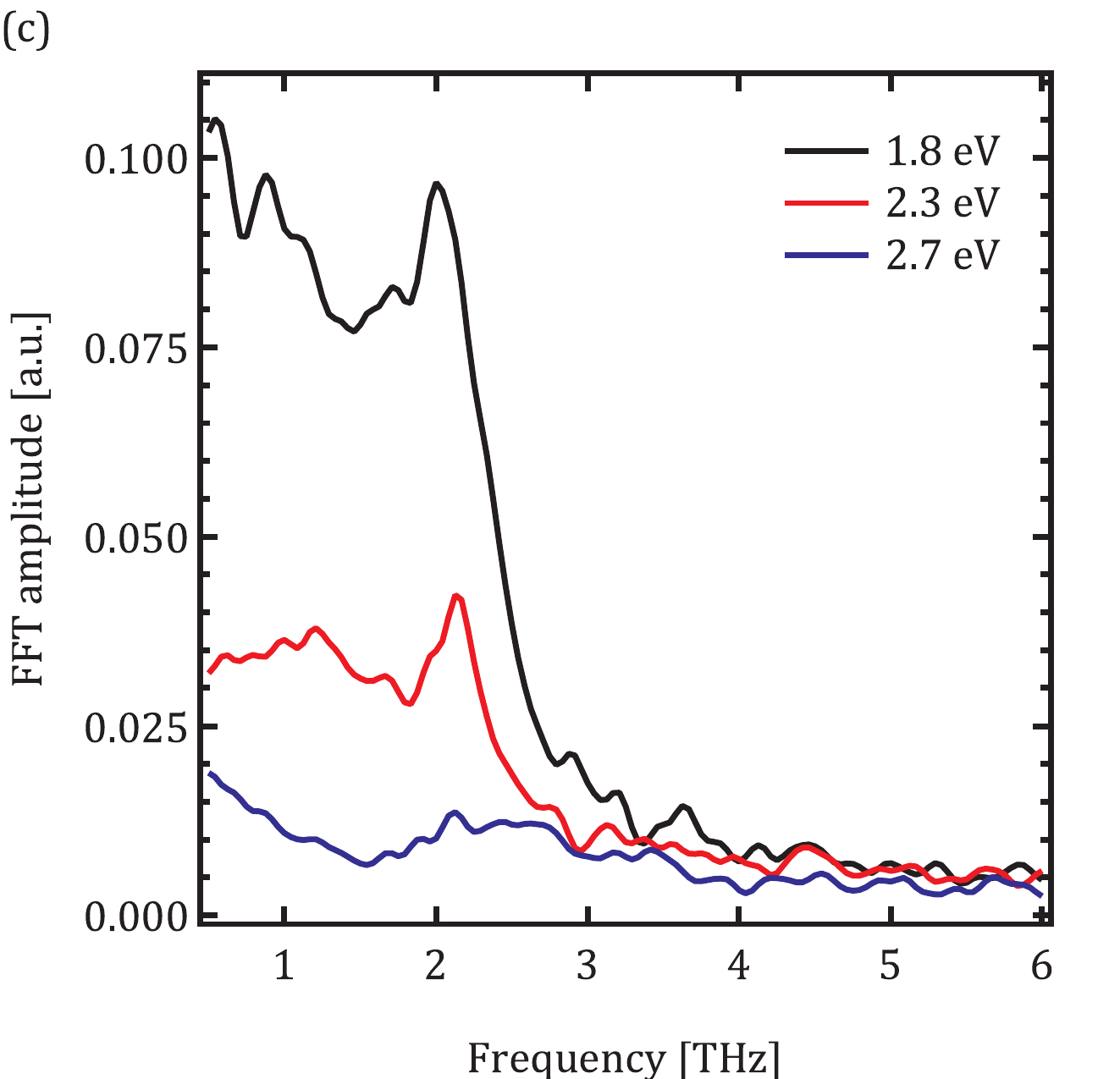}
	\caption[Fourier transforms for TaS2]{
	Fourier transform of the traces from Fig.~\ref{fig:Fig1}(d-f) at (a) 10~K, (b) 230~K, and (c) 300~K, revealing the coherent oscillations at 2.4~THz resp. 2.1~THz and at 3.8~THz resp. 3.6~THz. 
	The additional (asymmetric) mode at 2.1~THz at 10~K is not distinguishable from the transform. 
	}
	\label{fig:SIFig1}
\end{figure}
Figure~\ref{fig:SIFig1} displays the Fourier transforms of the traces from Fig.~\ref{fig:Fig1}(d-f) for all measured temperatures. 
At 10~K, two out of the three appearing oscillations can be identified, while the third one cannot be separated from the amplitude mode peak. 
At higher temperatures, the oscillations are much harder to discern.

\subsection{Full data sets for the fluence dependence at 10~K}
\begin{figure*}[t]
	\includegraphics[width=.33\linewidth]{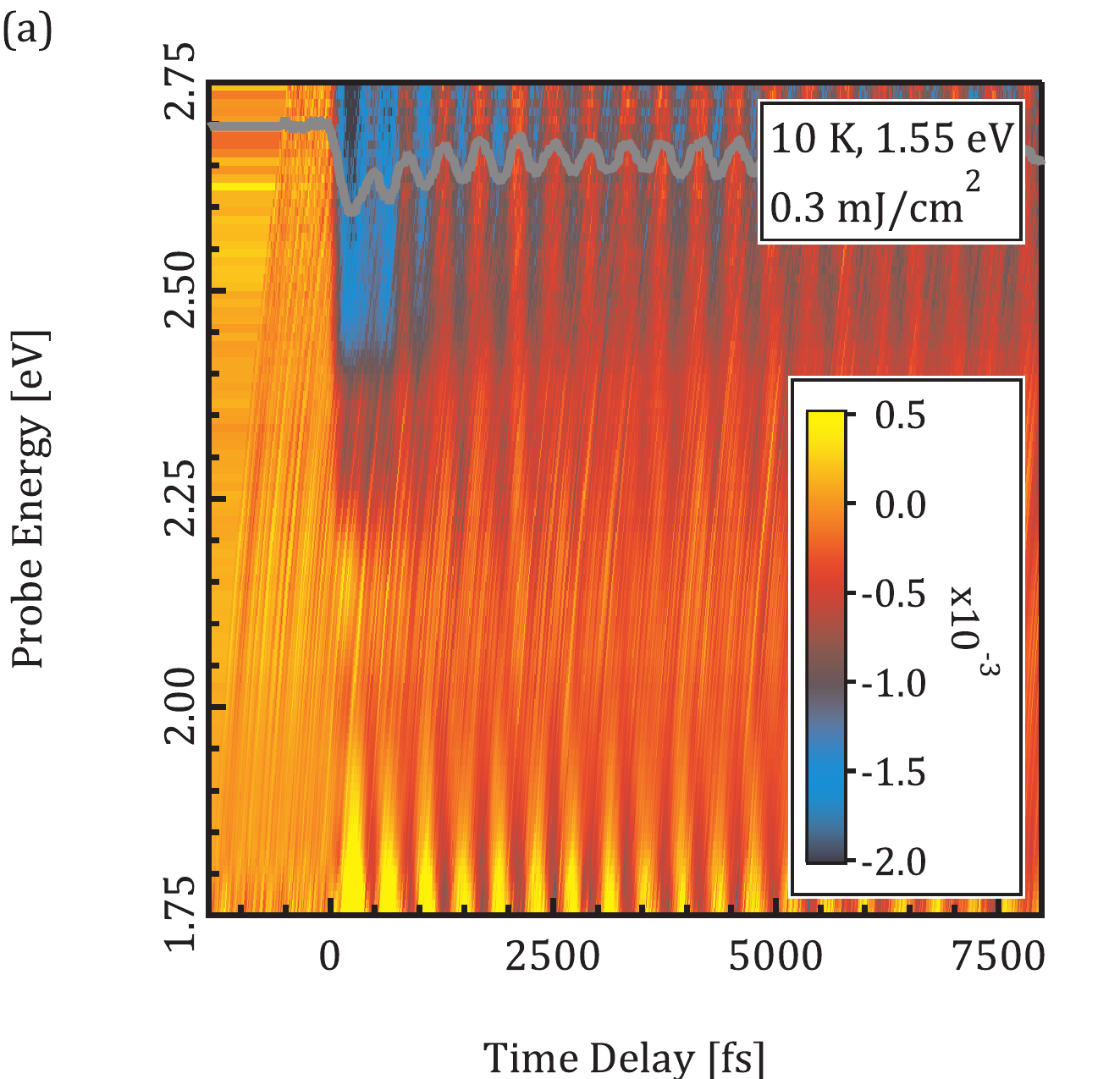}\hfill
	\includegraphics[width=.33\linewidth]{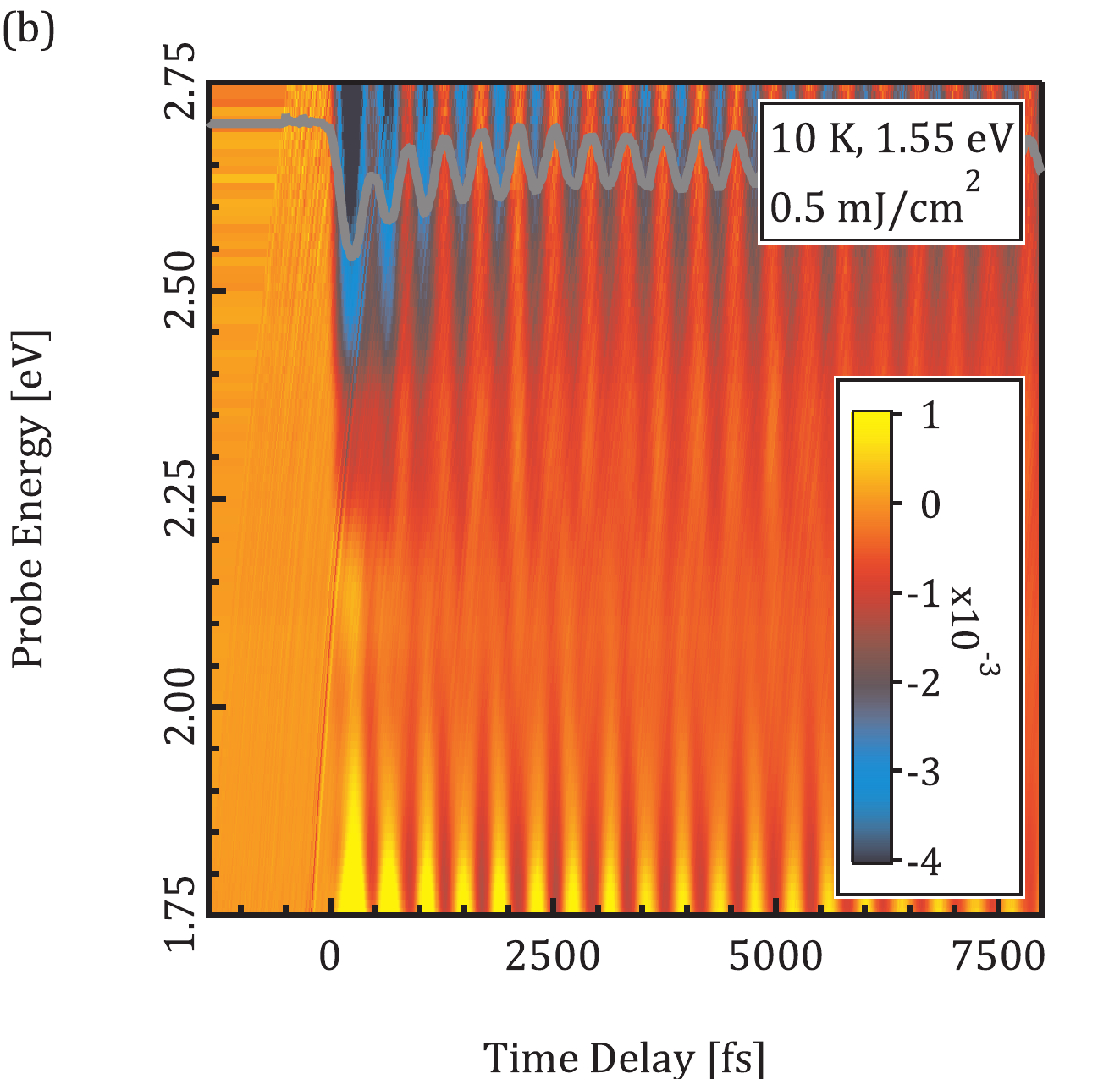}\hfill
	\includegraphics[width=.33\linewidth]{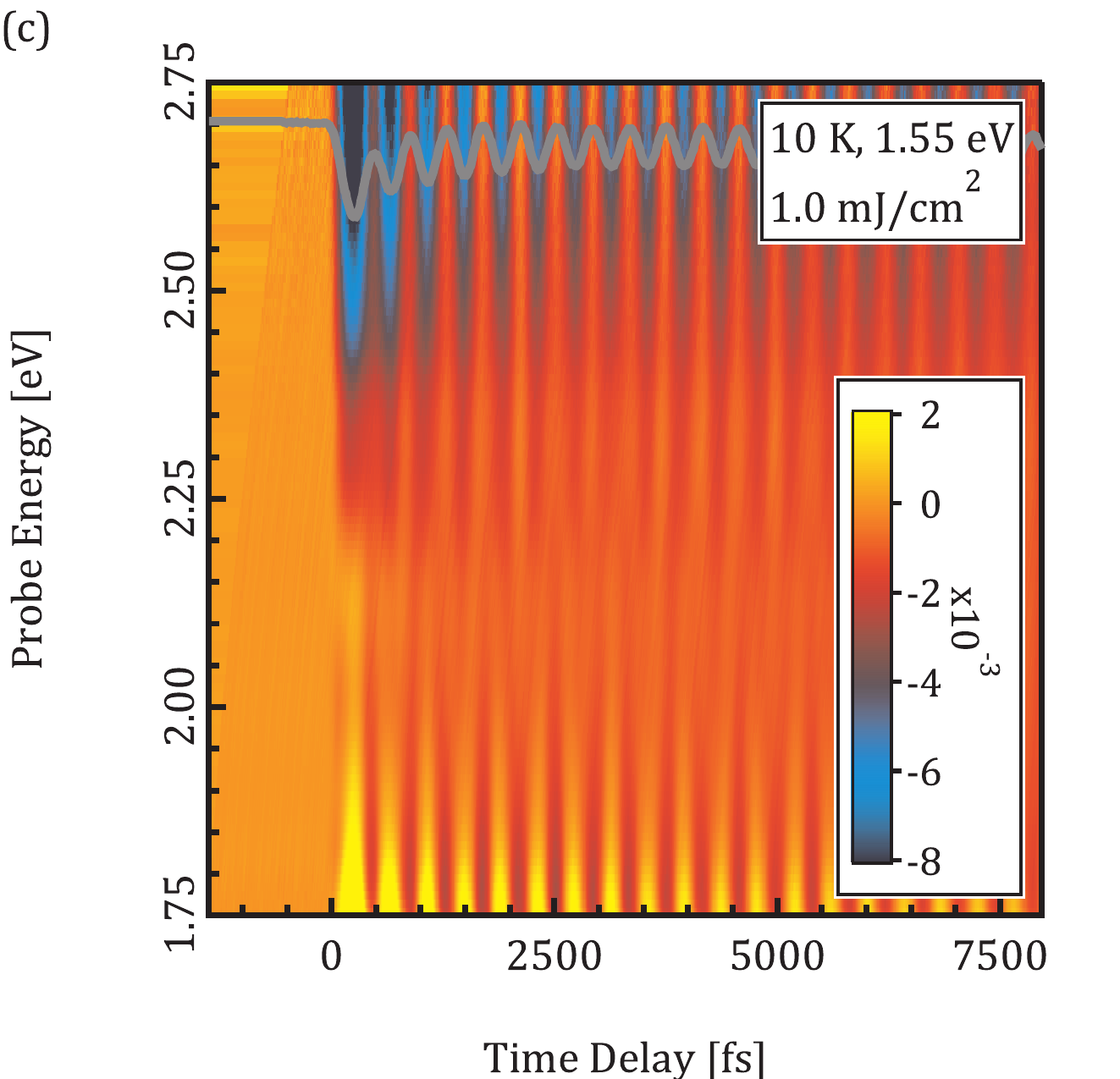}\\
	\includegraphics[width=.33\linewidth]{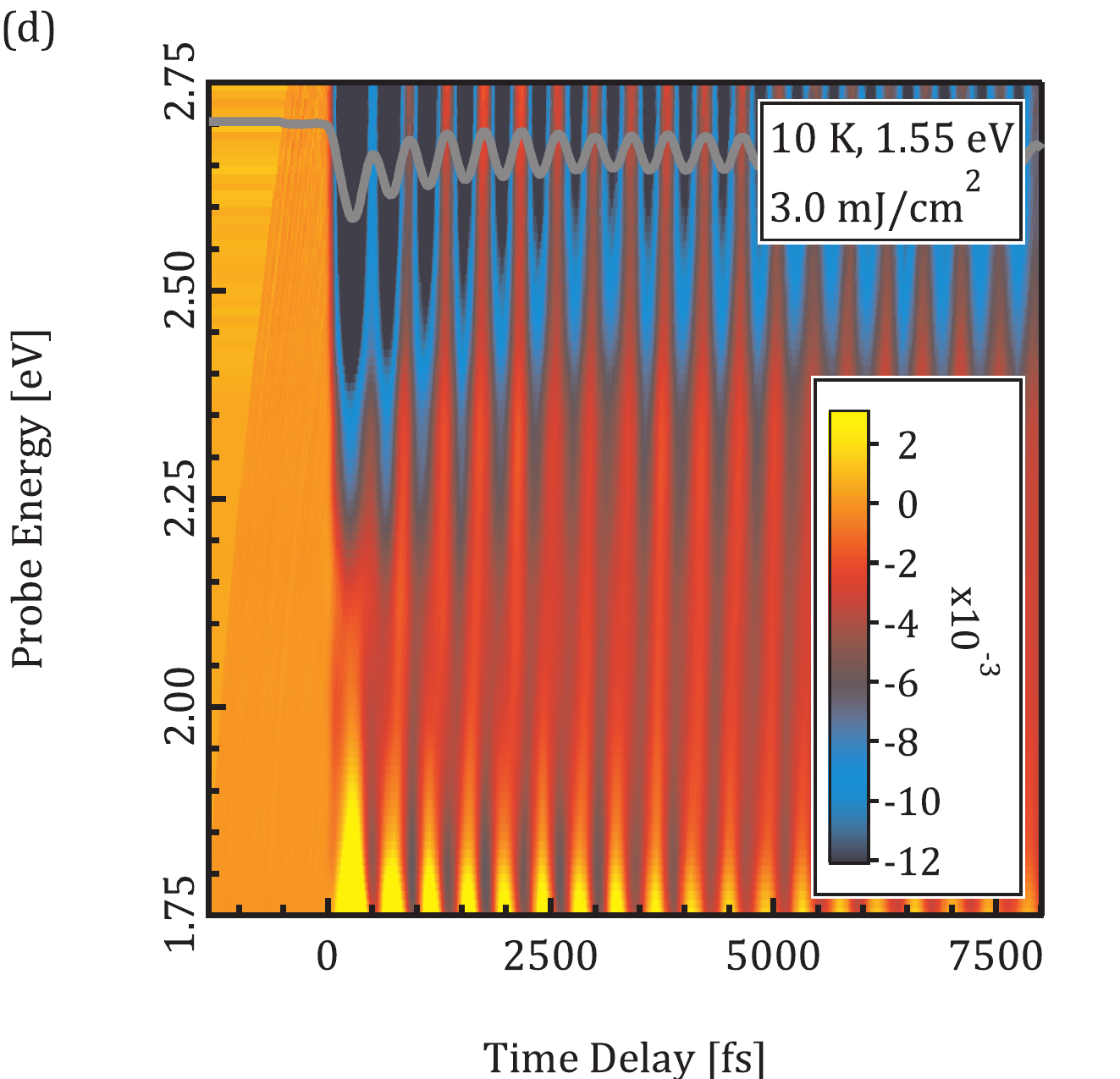}\hfill
	\includegraphics[width=.33\linewidth]{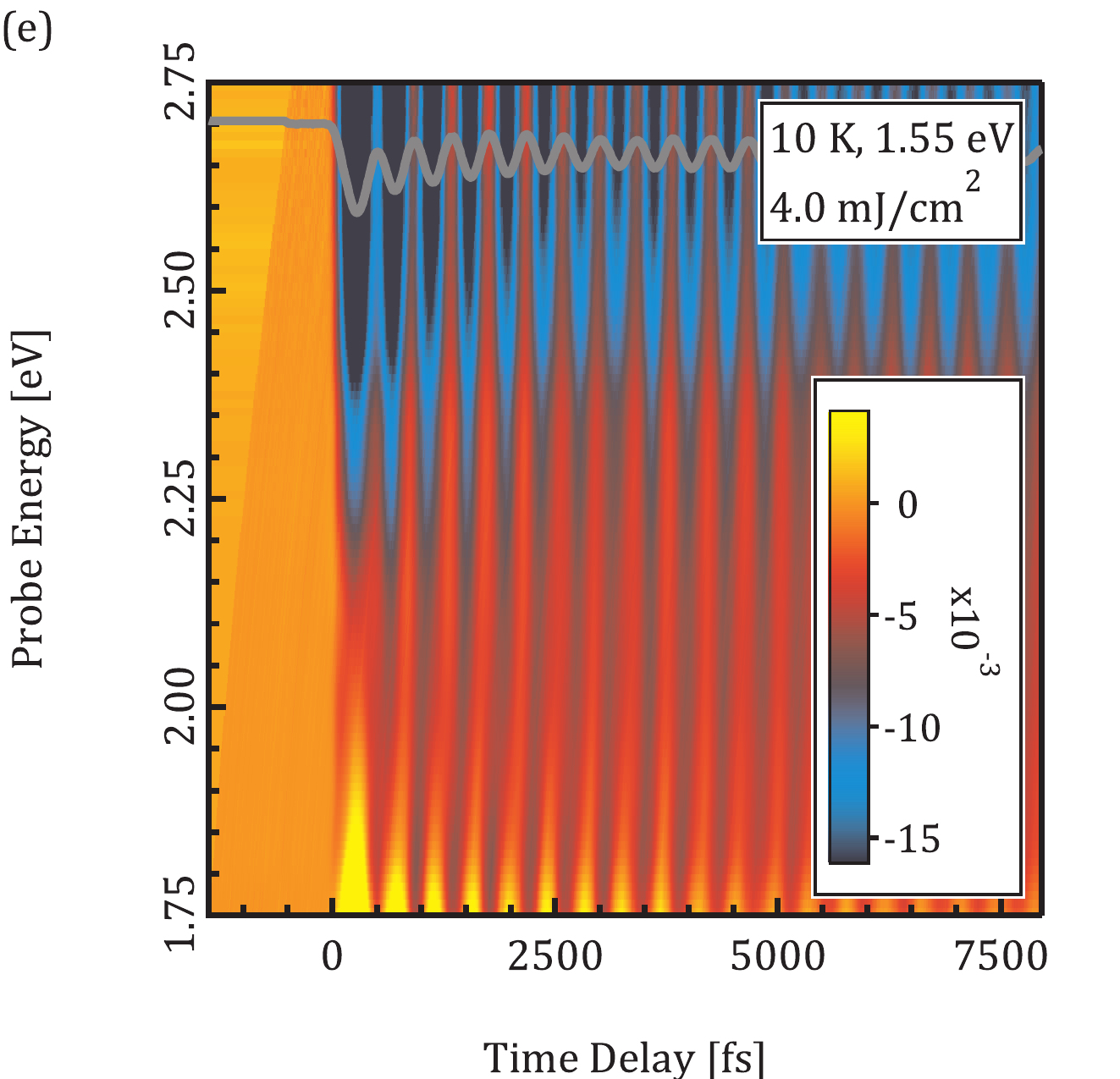}\hfill
	\includegraphics[width=.33\linewidth]{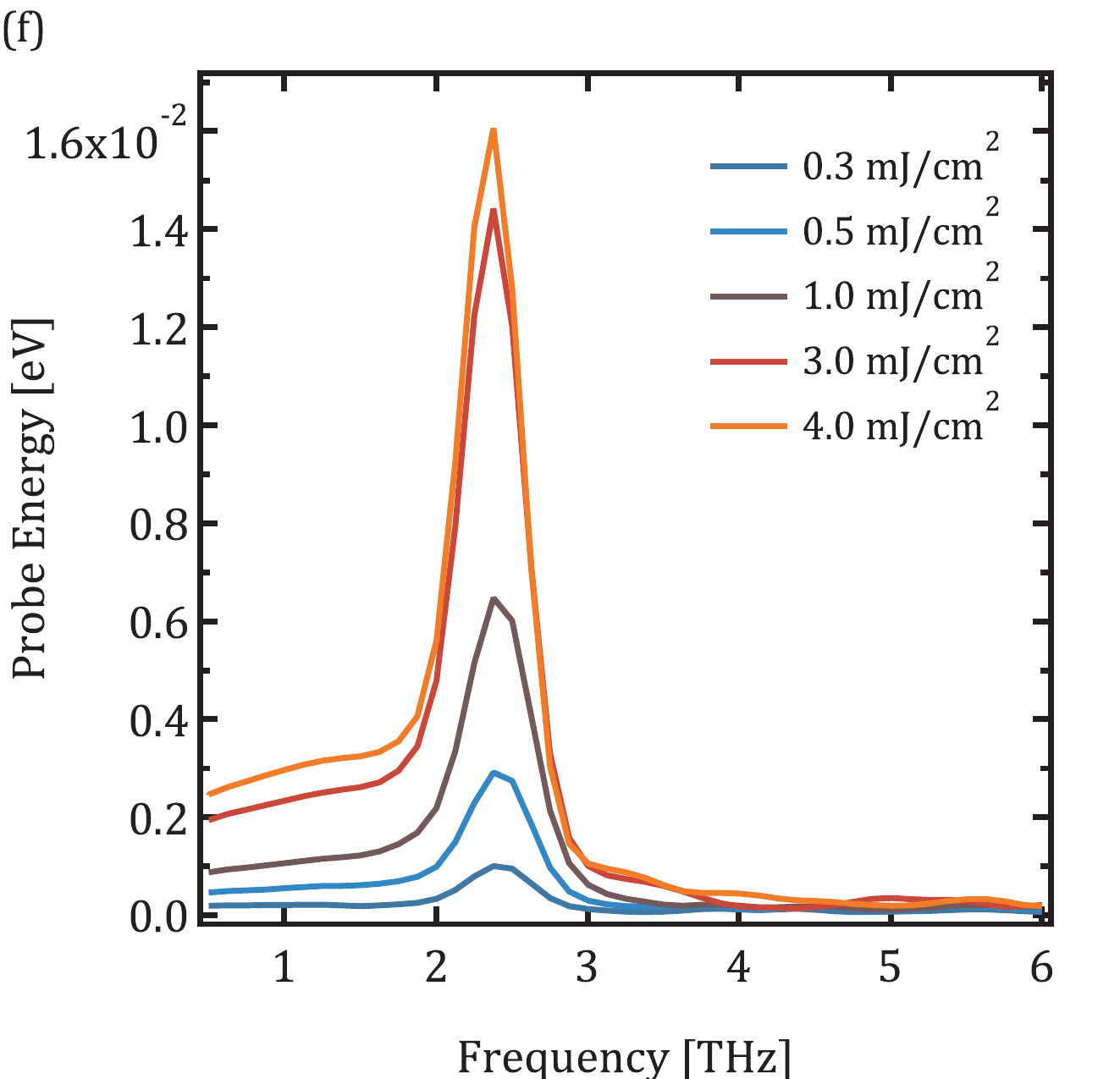}
	\caption[Fluence dependence for TaS2 at 10~K for 1.55~eV excitation energy]{
	Fluence dependence of the transient reflectivity change of TaS$_2$ at 10~K for 1.55~eV excitation energy. 
	The absorbed fluence is indicated in the labels. 
	The overlaid traces show the time evolution of the signal at 2.7~eV probe energy. 
	The last panel shows the Fourier transform of the overlaid traces. 
	}
	\label{fig:SIFig2}
\end{figure*}
Figure~\ref{fig:SIFig2} shows the fluence dependence of the transient reflectivity at 10~K, in the c-CDW phase, for excitation at 1.55~eV. 
The spectral features remain similar over a fluence range of an order of magnitude, except for the slight saturation effect and redshift of the amplitude mode zero-crossing discussed in the main text. 
The time traces at a probe energy of 2.7~eV, close to the maximum signal strength, are overlaid in the spectra, and are representative of the general trend: 
In the last panel of the figure, the Fourier transforms of the traces at 2.7~eV are shown. 
Note that the 3.8~THz mode is barely visible at 2.7~eV probe energy.

\subsection{Estimate of the maximum sample temperature increase at 300~K}
\begin{figure*}[t]
	\includegraphics[width=.5\linewidth]{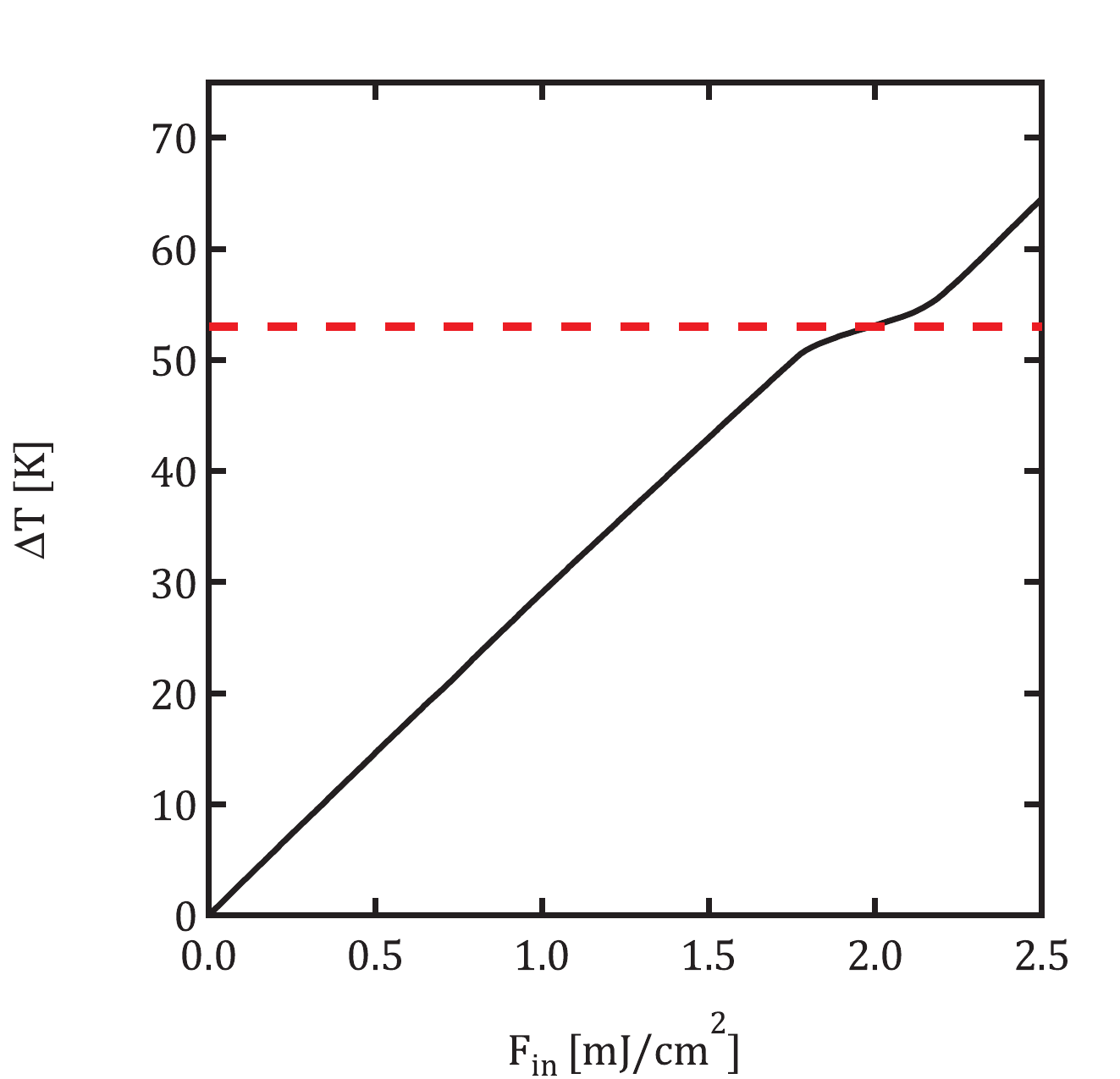}
	\caption{Estimate of the maximum lattice temperature increase $\Delta T$ as a function of incident fluence at 300~K. 
	The broken line indicates the transition to the i-CDW phase. }
	\label{fig:SIFig3}
\end{figure*}
We estimate the maximum sample temperature increase $\Delta T$ based on heat capacity data reported by Suzuki et al.~\cite{Suzuki1985}, given in cal/(mol$\cdot$K).
For simplicity, we assume homogeneous illumination of a cylindrical volume $V = A\cdot \delta$, where $A=(150\mu\text{m})^2$ is the illuminated area assuming a top-hat profile, and $\delta = 70$~nm is the penetration depth at 1.55~eV (cf. Fig.~\ref{fig:Fig3}(c) of the main text). 
Using a molar mass $M = 245.08$g/mol and a density $\rho_M = 6.86$g/cm$^3$, we find the illuminated amount of substance to be
\[
	n = \frac{V\cdot \rho_M}{M} = 4.4\cdot 10^{-11}~\text{mol}. 
\]
The maximum $\Delta T$, reached directly after equilibration of electrons and lattice, can be estimated via the implicit relation
\[
	W_{dep} = n\cdot C_p^{tot}(\Delta T),
\]
where $W_{dep} = (1-R)\cdot F_{in}\cdot A$ is the deposited energy, which is proportional to the incident fluence $F_{in}$, and the total heat capacity $C_p^{tot}$ is given by the integral
\[
	C_p^{tot} = \int_{T_0}^{T_0+\Delta T}\limits C_p(T')dT'. 
\]
Using a starting value of $T_0 = 300$~K, we plot $\Delta T$ as a function of the incident fluence in Fig.~\ref{fig:SIFig3}. 
We point out that in order for the sample to enter the i-CDW phase for a finite amount of time an incident fluence greater than 2~mJ/cm$^2$ has to be chosen.

\subsection{Full data sets for excitation in the near-IR}
\begin{figure*}[t]
	\includegraphics[width=.33\linewidth]{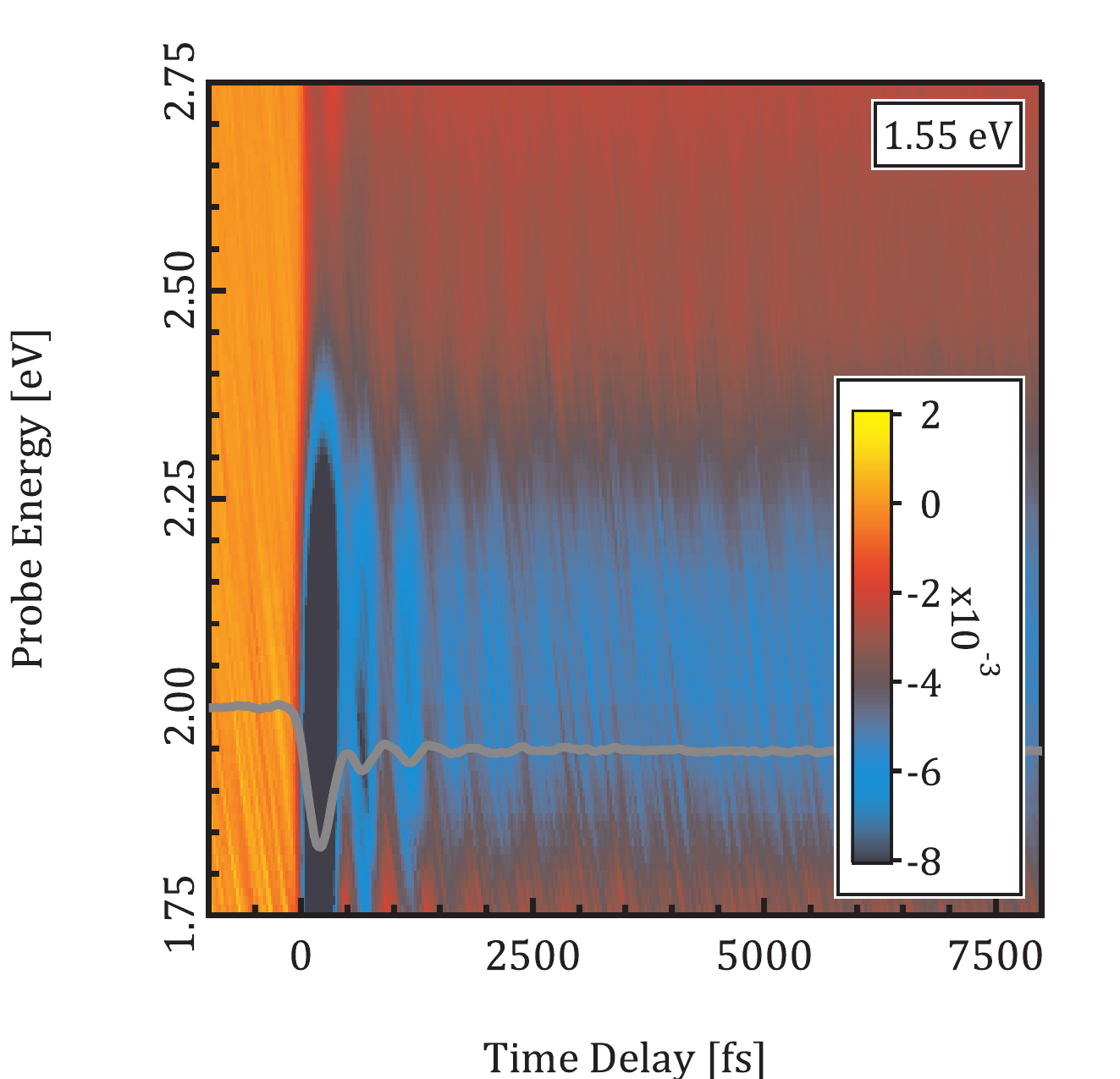}\hfill
	\includegraphics[width=.33\linewidth]{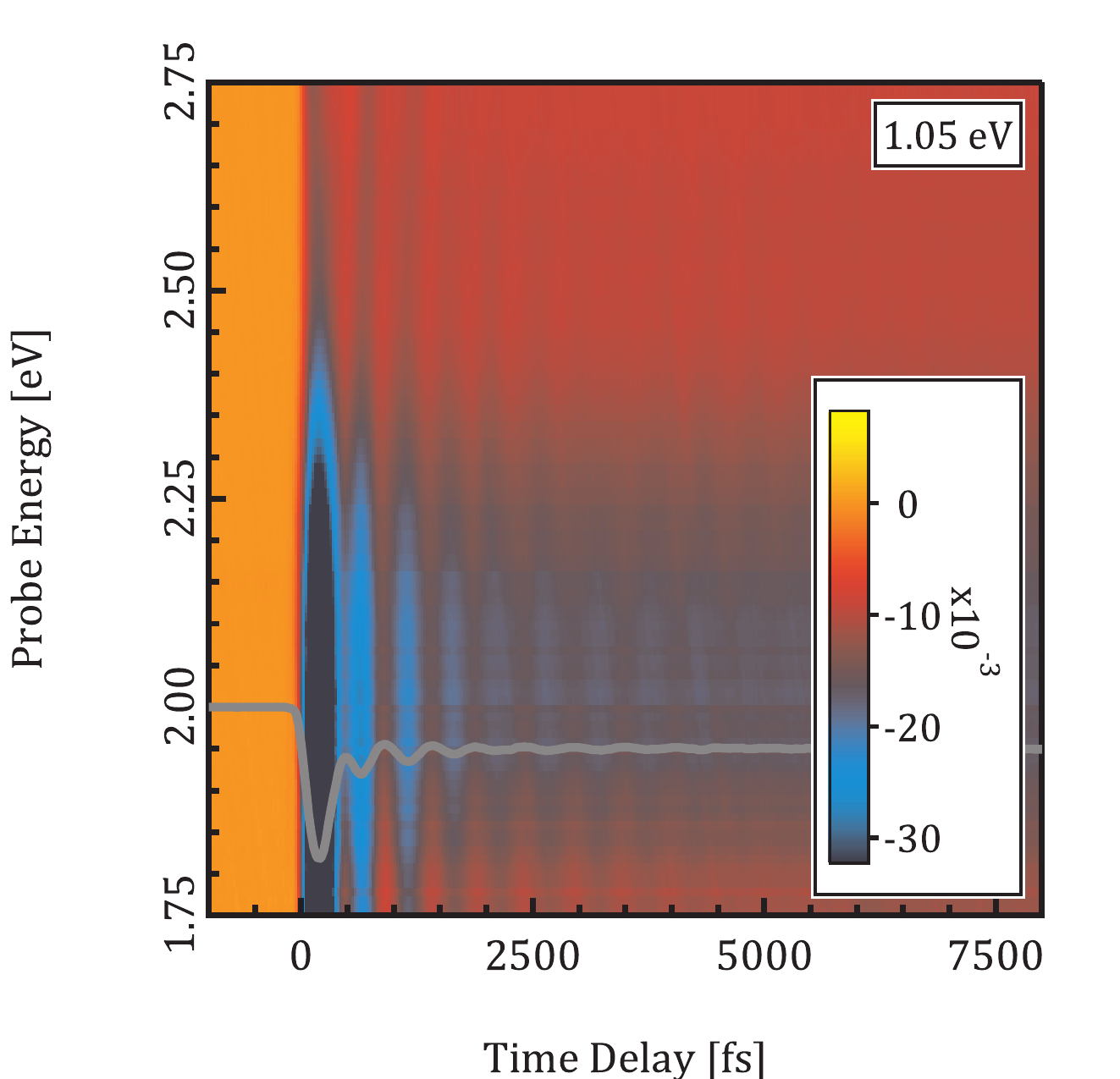}\hfill
	\includegraphics[width=.33\linewidth]{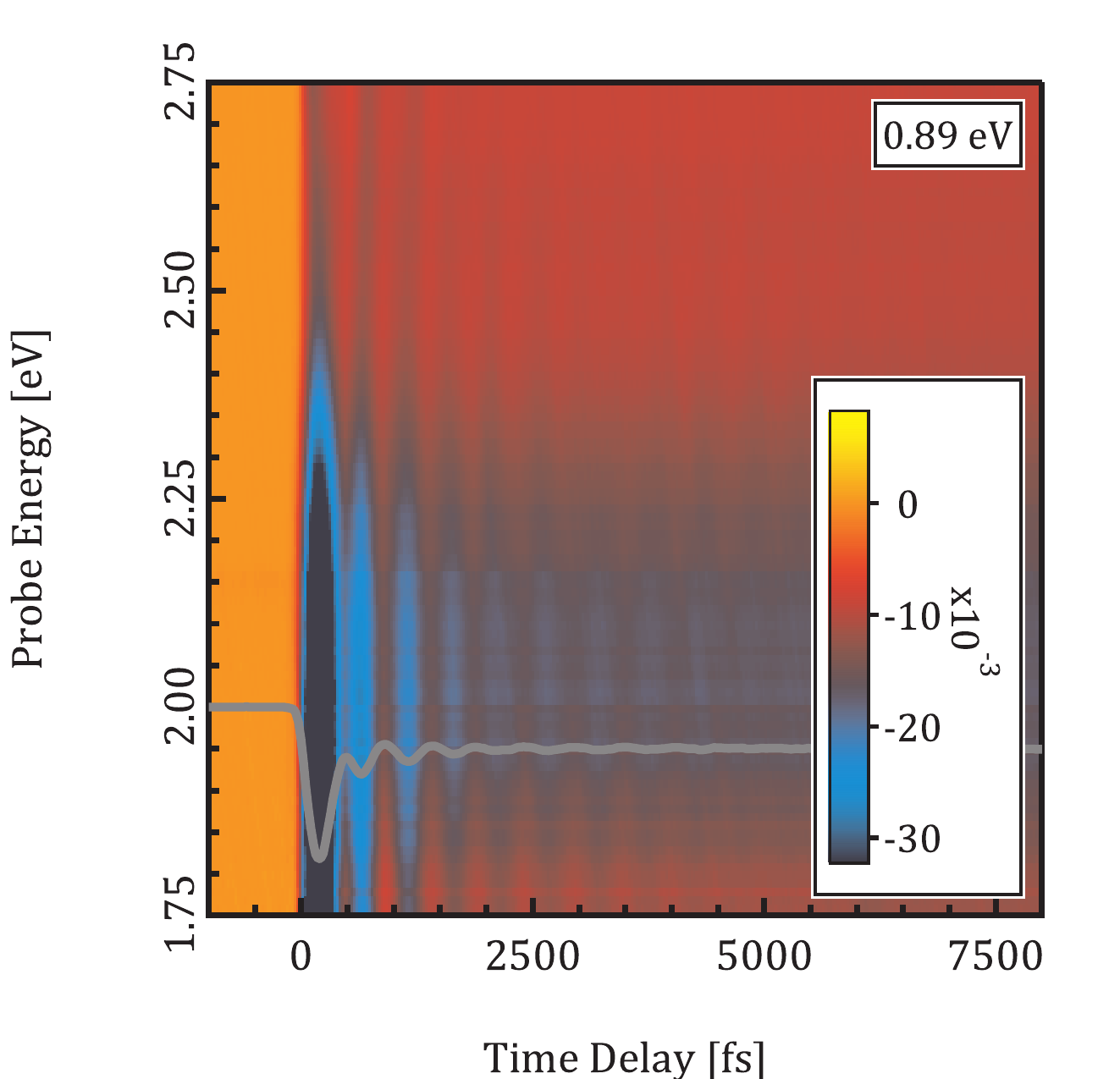}\\
	\includegraphics[width=.33\linewidth]{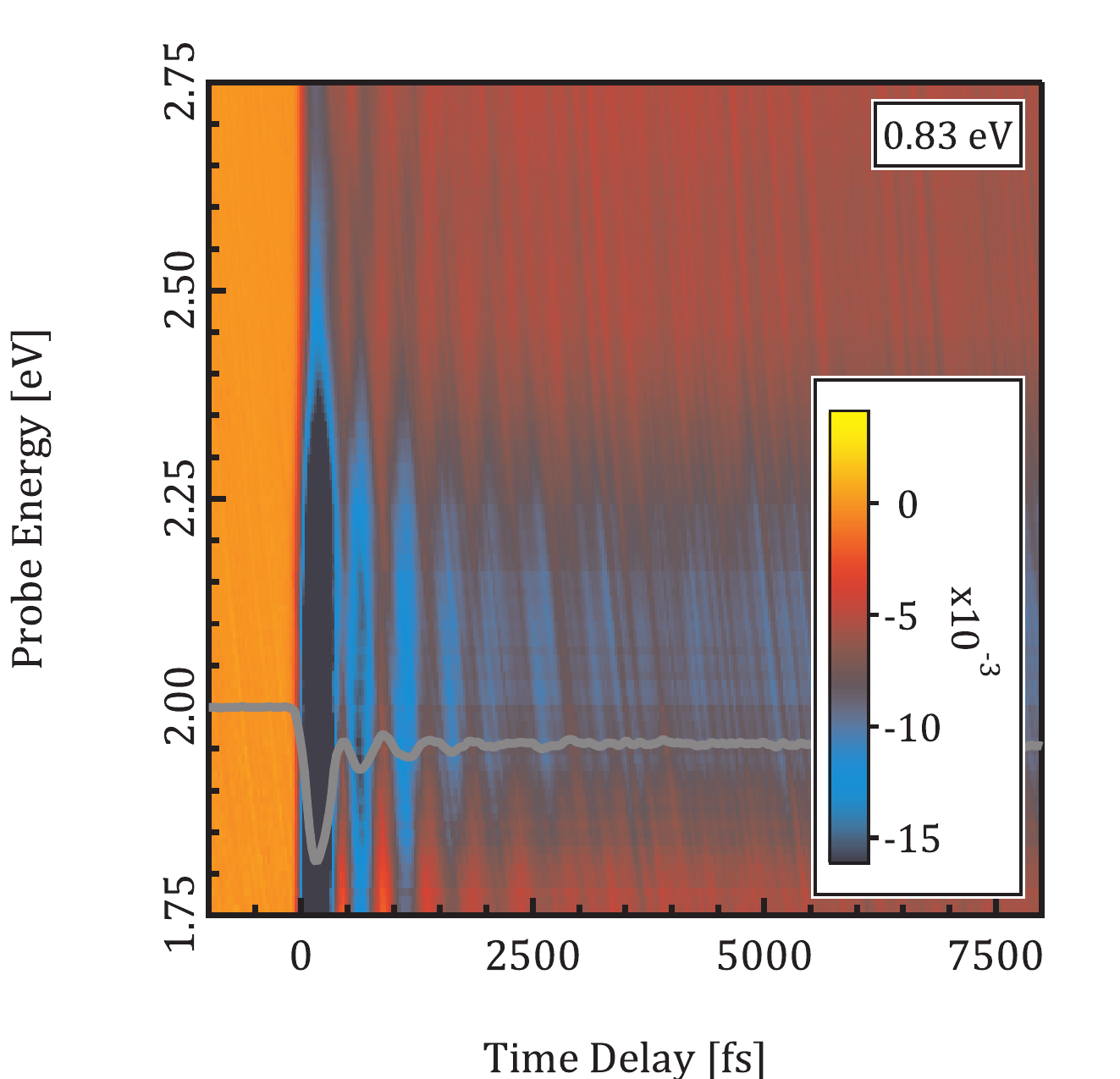}\hfill
	\includegraphics[width=.33\linewidth]{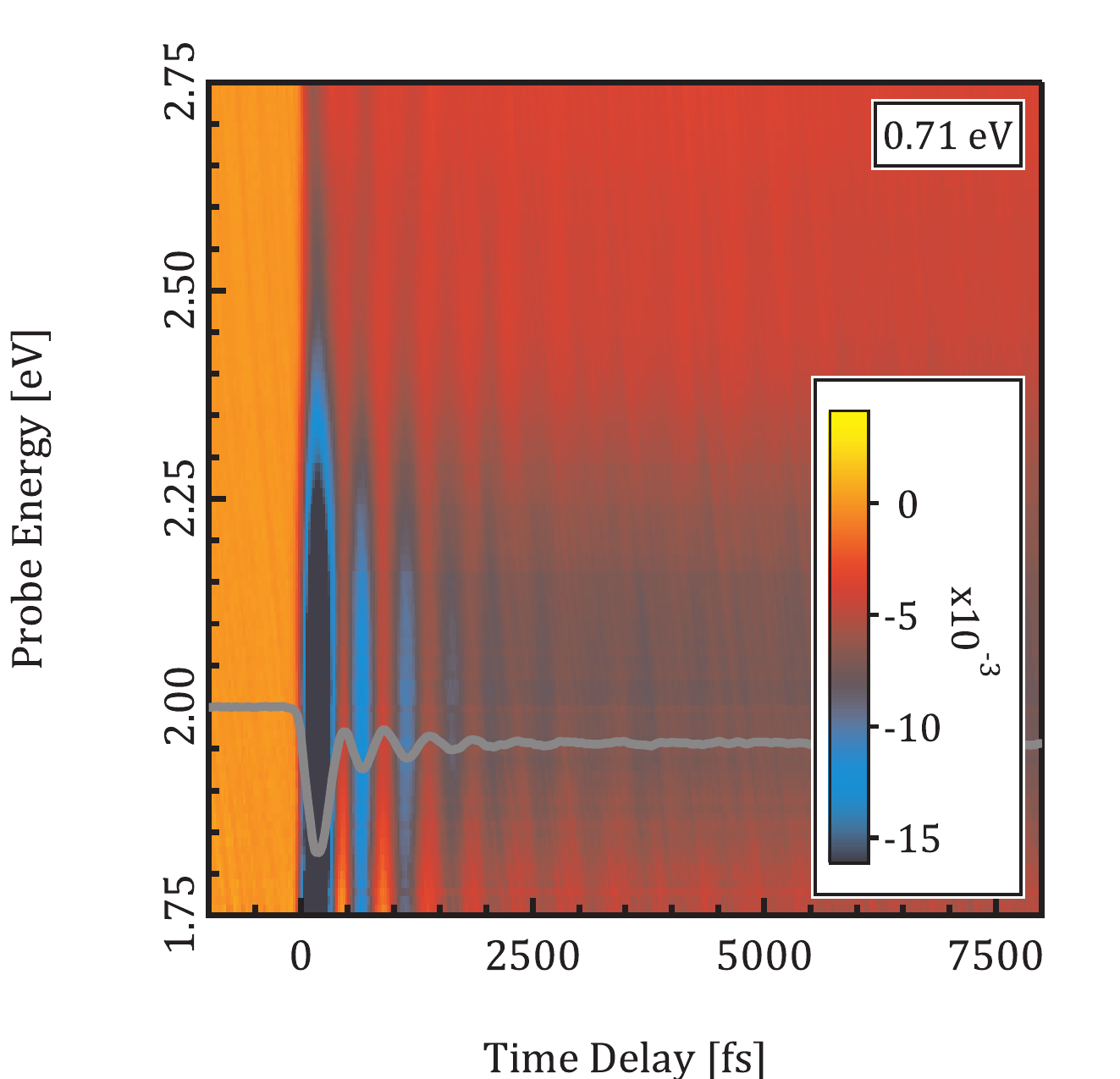}\hfill
	\includegraphics[width=.33\linewidth]{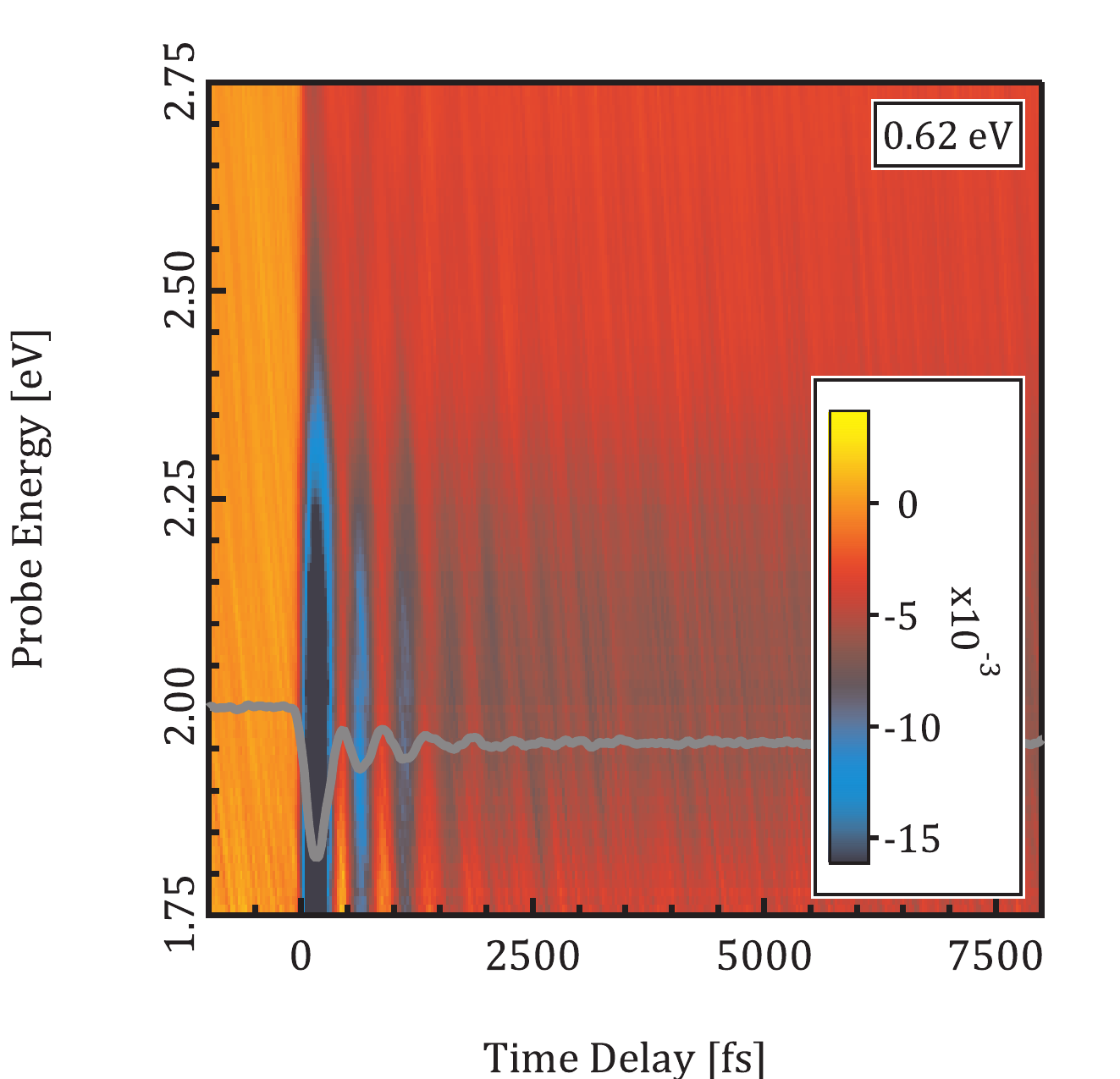}\\
	\includegraphics[width=.33\linewidth]{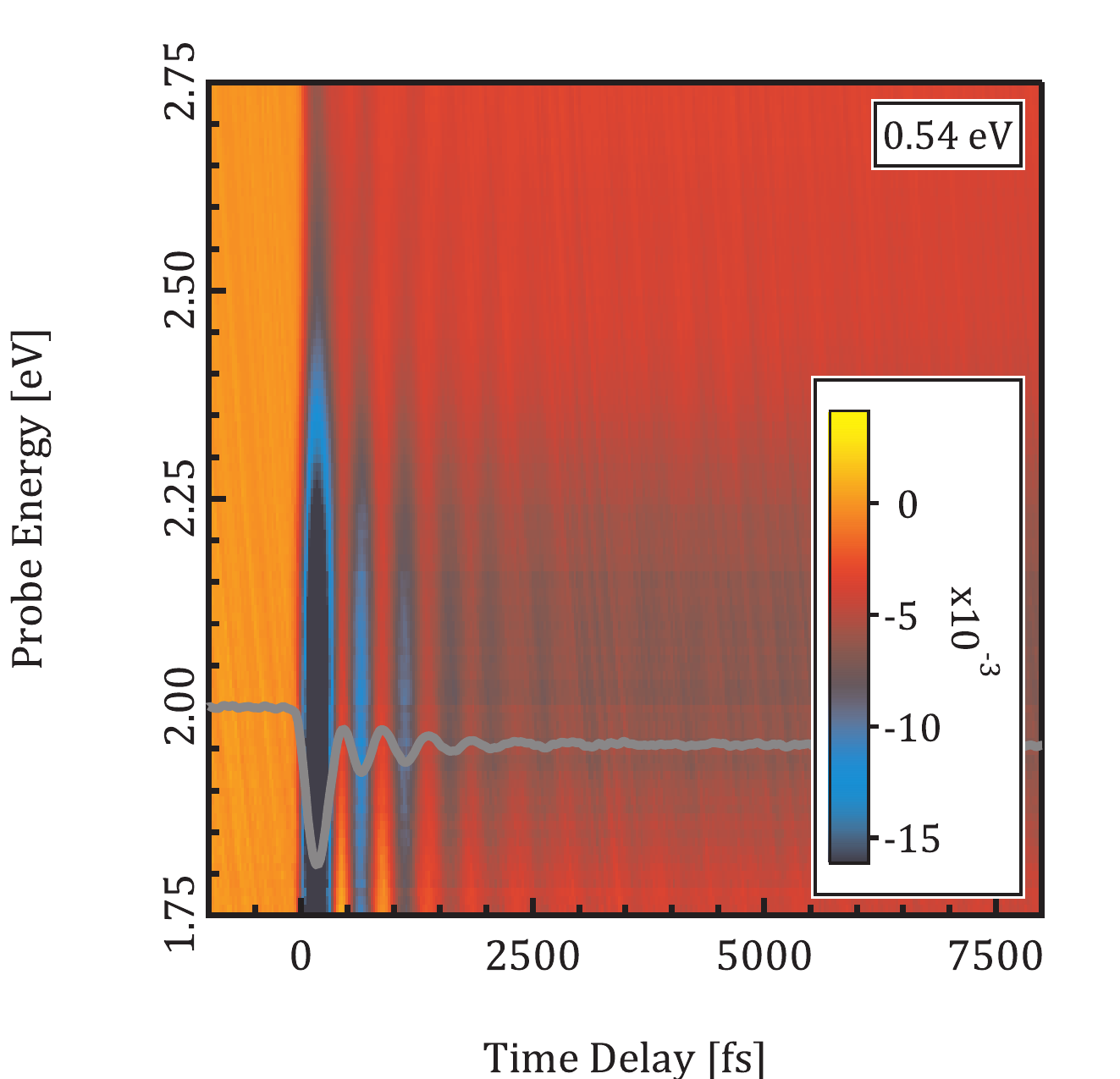}\hfill
	\hspace{.33\linewidth}\hfill
	\hspace{.33\linewidth}
	\caption[Excitation energy dependence for TaS2 at 300~K]{
	Excitation energy dependence of the transient reflectivity change of TaS$_2$ at 300~K in the near-IR. 
	The excitation energy is indicated in the labels. 
	The overlaid traces show the time evolution of the signal at 2~eV probe energy. 
	}
	\label{fig:SIFig4}
\end{figure*}
Figure~\ref{fig:SIFig4} reports the transient reflectivity spectra in the n-CDW phase at 300~K for various excitation energies. 
The spectra all follow the general trend observed for the measurement at 300~K in Fig.~\ref{fig:Fig1}. 
The signal peaks around 2~eV probe energy, where the time traces are overlaid to the spectra, and is dominated by the incoherent background. 
A coherent oscillation at 2.1~THz is present and decays with approximately 1~ps. 
From the traces at 2~eV it can also be seen that the strength of the amplitude increases slightly when reducing the excitation energy.

\subsection{Correction for the penetration depth mismatch}
The pump beam creates an excitation profile that is assumed to be proportional to its intensity $I^{pu}$ as a funciton of the depth $z$,
\[
	I^{pu}(z) = I_0^{pu}\cdot e^{-z/\delta_{pu}}, 
\]
where $\delta_{pu}$ is the penetration depth of the pump. 
The measured total differential reflectivity $\Delta R(\omega)$ can be expressed as an integral of the contributions $\Delta r(z,\omega)$ that are created at different depths $z$ and travel back to the surface to form the specular reflection. 
Each contribution $\Delta r(z,\omega)$ is assumed to be proportional to the intensity of the pump beam as well as the intensity of the probe beam at a given depth $z$: 
\[
	\Delta r(z,\omega) \propto I^{pu}(z)\cdot I^{pr}(z,\omega),
\]
where
\[
	I^{pr}(z,\omega) = I_0^{pr}\cdot e^{-z/\delta_{pr}}. 
\]
The total differential reflectivity is now given as
\[
	\Delta R(\omega) \propto \int_0^\infty\limits \Delta r(z,\omega)\cdot e^{-z/\delta_{pr}} dz. 
\]
Here, the additional factor $e^{-z/\delta_{pr}}$ represents the attenuation experienced by the backreflected probe light before leaving the sample.

Altogether, we have
\begin{equation}
	\Delta R(\omega) = \int_0^\infty\limits I_0^{pu}\cdot I_0^{pr}\cdot e^{-z/\delta_{pu}}\cdot e^{-2z/\delta_{pr}} dz. 
	\label{eq:delta_r_meas}
\end{equation}
We define the ``real'' differential reflectivity $\Delta R^\ast(\omega)$ as the quantity that is obtained if the penetration depth of the probe beam is equal to that of the pump beam,
\begin{equation}
	\Delta R^\ast(\omega) = \int_0^\infty\limits I_0^{pu}\cdot I_0^{pr}\cdot e^{-z/\delta_{pu}}\cdot e^{-2z/\delta_{pu}} dz. 
	\label{eq:delta_r_real}
\end{equation}
Combining Eqs.~\eqref{eq:delta_r_meas} and~\eqref{eq:delta_r_real} leads to the correction
\[
 \Delta R^\ast(\omega) = \Delta R(\omega)\cdot \frac{2/\delta_{pr}+1/\delta_{pu}}{3/\delta_{pr}}. 
\]

\end{document}